\documentclass{easychair}
\usepackage[utf8]{inputenc}
\usepackage[normalem]{ulem}
\usepackage{stmaryrd,amssymb, amsmath,mathtools,amsthm,graphicx}
\usepackage{color}
\usepackage[binary-units=true,group-four-digits=true,detect-all]{siunitx}
\usepackage{wasysym,xspace}
\usepackage{url,pifont,cite,doi}
\usepackage{pdfpages}
\usepackage{pifont}
\usepackage{multirow,booktabs}
\usepackage{subcaption,enumitem,etoolbox,nicefrac}
\usepackage{bold-extra} 
\usepackage[noend]{algorithm2e} \SetAlCapSkip{1em}

\usepackage[final,tracking=true,kerning=true,spacing=true,stretch=10,shrink=10,babel=true,letterspace=25]{microtype}
\microtypecontext{spacing=nonfrench}

\usepackage[english]{babel}
\makeatletter\AtBeginDocument{\let\@elt\relax}\makeatother
\hypersetup{
  pdftex,
  pdftitle={Scalable Proof Producing Multi-Threaded SAT Solving with Gimsatul through Sharing instead of Copying Clauses},
  pdfauthor={Mathias Fleury and Armin Biere},
  pdfkeywords={parallel SAT solving, proofs, gimsatul, clause sharing},
  pdfborder={0 0 0},
  bookmarksnumbered,
  bookmarksdepth=3,
  bookmarksopenlevel=3,
  bookmarksopen}

\iftrue
\makeatletter

\def\orcid#1{\smash{\href{http://orcid.org/#1}{\protect\raisebox{-1.25pt}{\protect\includegraphics{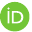}}}}}

\makeatother
\fi

\newif\ifcompactfigures
\compactfiguresfalse

\newcommand{\tools}[1]{\textsc{#1}}

\newcommand{\kissat}{\tools{Kissat}\xspace}
\newcommand{\glucose}{\tools{Glucose}\xspace}

\newcommand{\gimsatul}{\tools{Gimsatul}\xspace}
\newcommand{\zchaff}{\tools{zChaff}\xspace}
\newcommand{\manysat}{\tools{ManySAT}\xspace}

\newcommand{\pmcomsps}{\tools{P-Mcomsps}\xspace}

\iftrue
\newcommand{\PARAGRAPH}[1]{\par\smallskip\noindent\textbf{#1}~\nolinebreak}
\else
\newcommand{\PARAGRAPH}[1]{\paragraph{#1}}
\fi

\DeclareSIUnit\loc{loc}

\begin{document}
\title{Scalable Proof-Producing Multi-Threaded SAT Solving \\
with Gimsatul through Sharing instead of Copying Clauses}
\titlerunning{Clause Sharing in Parallel SAT solving}
\author{Mathias Fleury~\orcid{0000-0002-1705-3083} and Armin Biere~\orcid{0000-0001-7170-9242}}
\authorrunning{Mathias Fleury and Armin Biere}
\institute{
  Albert-Ludwig-Universität Freiburg, Germany \\
  \email{fleury@cs.uni-freiburg.de \quad biere@cs.uni-freiburg.de}
}

\maketitle

\begin{abstract}
We give a first account of our new parallel SAT solver Gimsatul.  Its key
feature is to share clauses physically in memory instead of copying them,
which is the method of other state-of-the-art multi-threaded SAT solvers to
exchange clauses logically.  Our approach keeps information about which
literals are watched in a clause local to a solving thread but shares the
actual immutable literals of a clause globally among all solving threads.
This design gives quite remarkable parallel scalability, allows aggressive
clause sharing while keeping memory usage low and produces more compact proofs.
\end{abstract}

\section{Introduction}
\label{sec:introduction}

The SAT Competition is the place to show off the newest and fastest SAT solvers
for many years now. To improve reliability and increase correctness, every
solver in the main track must produce a DRAT certificate that is checked by the
official checker \tools{drat-trim} -- and only checked problems count as solved.
Solvers implementing techniques that cannot be expressed in DRAT can only take
part in the no-limit track (even if no technique seems to bring an edge over
other main-track solvers).
While sequential solvers have improved considerably recently, making use of multiple CPU cores can
improve performance even further. Therefore the SAT Competition has a parallel
and a cloud track. In those track no proofs are required.

There are several ways to parallelize SAT
solvers~(Section~\ref{sec:parallel-sat-solving}), but most solvers in both the
parallel and the cloud track rely on the common portfolio approach pioneered
by \manysat~\cite{hamadi2008manysat} enhanced by exchanging clauses. The idea is to rely
on existing fast single-core SAT solvers without the need to modify them beyond possibly
adding a mechanism for exchanging clauses.  The main approach to exchange
clauses between solver threads is copying them.

The main argument in favor of this approach is that adding a
light-weight synchronization to existing code is simple and does not need to
change the sophisticated and complex core data-structures of the solvers.  We argue that this
argument leads to sub-optimal results both with respect to scalability and
memory usage. Instead we propose to physically share clause data (through
pointers) during clause exchanges instead of copying them.

More precisely, the classical portfolio approach
generates two issues (Section~\ref{sec:proof-check-relat}). First, it increases the amount of memory needed
by each solver thread. Second, even if proof production was implemented, it produces larger proofs than necessary by requiring
duplicating shared clauses in the proof.
By actually sharing instead of copying clauses we can address both issues.

Current proof checkers do not support exchanging clauses
between certificates and it is not obvious how to do this correctly, because
clause exchanges go in both directions. The other issue is that proof checking cannot
completely independently be done in parallel, again, because the solver
threads exchange clauses.
One partial solution to the problem is to log all the clauses into a single file and have a
global clock to serialize all the derived clauses in order.  However this
solution still requires to duplicate exchanged clauses in the proof log, leading to
large traces and long checking times, as also our experiments confirm.

In order to reduce this work, we have started a new SAT solver called
\gimsatul that aggressively shares clauses among the solver instances.  We
revisit an old idea that co-existed with the currently dominating
exchange-only approach until around 2011, which on the other hand pre-dates more sophisticated
proof tracing techniques used in the SAT competition now as well as all the improvements
to sequential solving made in the last decade.   We observe that physically
sharing clauses not only
makes it possible to reduce the memory footprint, but also enables sharing
in the proof log, thus reducing the amount of work required for checking.
While these changes have a large impact on the core data-structures of
the SAT solver they do not require any change to existing proof checkers.

We implemented our new solver \gimsatul in \SI{13}{\kilo\loc} of C. It is
available online\footnote{\url{https://github.com/arminbiere/gimsatul}} and
uses atomic operations to adjust reference counters and exchange pointers,
as standardized with C11 in \texttt{stdatomic.h}. It relies on the
Pthreads programming model for threads, locking, and condition variables.
Furthermore it uses lockless fast-path code whenever reasonable.

The name \gimsatul is derived from \emph{gimbatul} in the ``Black
Speech'' language invented by R.~Tolkien and occurs in the inscription of
the ``One Ring'' in ``Lord of the Rings'' and literally translates to ``find
them'' (all).  We follow that terminology in the paper and in the source
code.  Accordingly the main thread which performs preprocessing sequentially and
organizes everything is called the \emph{Ruler} and an actual solver thread
is called \emph{ring}.

This paper is a slightly extended version of our POS'22 paper made available to
attendees of the workshop on Pragmatics of SAT (POS'22).  In this paper we use additional space
to improve readability of figures by increasing their size.  We further include various comments
present in the original longer submission, which had to be removed due to the
page limit for the final version at the workshop of 14 pages (plus references).

Besides briefly going over the architecture of the solver we focus in this
paper on describing differences to the core data-structures compared to
other solvers.  In particular watched literals cannot be kept as
the first two literals in the clauses, since different solver instances may need to
watch different literals (Section~\ref{sec:sharing-clauses}).
The solver also performs sequential inprocessing, requiring to give back all
irredundant clauses to a single instance.
This new
infrastructure makes sharing possible and space efficient. The main
advantage as highlighted by our experiments is that our solver scales
\emph{linearly} with the number of threads. Another major consequence is that
sharing ensures that less memory is required to run \gimsatul (Section~\ref{sec:scale-experiments}).

Finally, we compare DRUP/DRAT proofs generated by \gimsatul{} (Section~\ref{sec:drat-proof-checking}) with the
version of the solver where sharing is not done at the proof level and
instead clauses are duplicated in the proofs
(Section~\ref{sec:experiments}). We show that proof size is largely independent
of the number of threads.

\section{Parallel SAT Solving and Related Work}
\label{sec:parallel-sat-solving}

This work does not attempt to define what SAT is and how SAT solvers works in
details. For those details, we refer to the \emph{Handbook of
Satisfiability}~\cite{BiereJarvisaloKiesl-SAT-Handbook-2021}. Detailed
knowledge about the inner working of SAT solvers is not required beyond the fact
that SAT solvers resolves clauses to derive new clauses and that those clauses can
be exchanged if two solvers work on the same problem. Non-satisfiability preserving
transformation are not done in parallel.

With respect to parallelization, we classify the
approaches used by SAT solvers into three different categories,
as shown in the following table:
\begin{center}
\begin{tabular}{l|ll|ll}
  &One solver&&Several solvers&\\
  \toprule
  One  & CDCL + simplification &(1) &  Portfolio &(2) \\
  Problem& (e.g., \tools{Kissat}~\cite{Biere-SAT-Competition-2021-solvers}) && (e.g., \tools{Mallob}~\cite{schreiber2021scalable})\\ \hline
  Multiple &Cube-and-conquer (CnC) by hand &(3) & CnC + resplitting + sharing  &(4) \\
  Problems& (e.g., Marijn Heule~\cite{marijnheule}) & &(e.g., \tools{Paracooba}~\cite{Biere-SAT-Competition-2021-solvers})\\
  \bottomrule
\end{tabular}
\end{center}
\noindent
Note that some
solvers also combine techniques like
\tools{Painless}~\cite{DBLP:conf/tacas/FriouxBSK19} from (2) and (4), but in
their default configuration from the SAT Competition, they use Approach (2).
In this work we reconsider Approach~(2) where current solvers
rely on mostly unmodified single-core SAT solver engines.
The current state-of-the-art solvers rely on (logically) \emph{exchanging}
clauses through copying but not physically sharing them.

Usually this approach further takes
advantage of the portfolio idea: different solver threads use
different strategies, e.g., different restart scheduling, decision
heuristics, etc.  The hope is that different instances learn clauses
useful in other threads, especially short clauses.

\iftrue
\gimsatul is far from being the first solver to use physical clause sharing. One
solver, \tools{SArTagnang}~\cite{Kaufmann11sartagnan} was discussed at the
Pragmatics of SAT workshop in 2011. Unfortunately, no performance discussion was
done to see how the solvers scales per thread. However, this solver was not
faster that clause exchanges. A more interesting difference is that they use one
thread to simplify the problem. Therefore, they have adapted the messages that
are exchanged: Instead of only exchange clauses, the message can also be that
the clause is subsuming another one which can be removed. An interesting
observation is that they save where the watch list was found last in one of
their configuration, which is similar to caching during search.

Other solvers like \tools{PaMiraXT}~\cite{DBLP:journals/jsat/SchubertLB09} use a
combination of cube-and-conquer and portfolio: The search space is initially
divided and each space is solved using several threads sharing clauses. On
motivation for the space splitting is that their implementation shares all
clauses (among the instance working on the same sub-problem). This is too much
for the poor SAT solver instances especially when 32 threads learn clauses at
the same time.

\fi

For detailed information on the architecture or the solver that used physical
sharing of clauses before 2011 (even though none of these solvers seems to be maintained
anymore) we refer to the corresponding chapter in the \emph{Handbook of Parallel Constraint Reasoning}~\cite{DBLP:books/sp/18/BalyoS18}.
Most current research tries to improve the portfolio approach and
investigates better
selection of clauses to exchange. Another interesting idea is to
let a GPU select the clauses which are useful~\cite{DBLP:conf/sat/PrevotSM21},
partially based on the idea the clauses that would have produced a conflict
earlier are likely useful in the future too. We
focus on scalability and proofs and  leave this aspect to future work.

\section{Proof Checking}
\label{sec:proof-check-relat}

In the previous section we discussed the different approach to solve problems.
Proof checking has a different flavor:
\begin{center}
\begin{tabular}{l|ll|ll}
  &One solver&&Several solvers&\\
  \toprule
  One  & one checker &(1) &  previous work and this paper &(2) \\
  Problem& (e.g., \tools{drat-trim}~\cite{DBLP:conf/sat/WetzlerHH14})& & (e.g., \tools{drat-trim}~\cite{DBLP:conf/sat/WetzlerHH14}) \\\hline
  Multiple &parallel checker& (3) & none \\
  Problems& (e.g., \tools{cake\_lpr}~\cite{DBLP:conf/tacas/TanHM21}) & \\
  \bottomrule
\end{tabular}
\end{center}
\noindent
Checking (1) is the most standard and best understood approach, even if there is
some technical issue on the semantics of (reused) units~\cite{DBLP:conf/sat/Rebola-PardoB18}.
The approach (3) is very promising to check cube-and-conquer proofs. The
checker checks each proof (i.e., the one with the cubes) and checks that the
cubes cover the entire search space. All those checks can be done in
parallel. There are some limitations in the verified proof checker of (3); for example,
cubes generated by \tools{March} cannot be used because trivially unsatisfiable cubes
are removed from the clauses, and more critically, the ``exchange'' of
information that the cubes are unsatisfiable is done via (forgeable) command line
arguments (while the checking itself is verified in CakeML).

In this paper we attempt to both improve the memory requirement by sharing physically
clauses and also, as a side effect, to reduce the size of the proofs.

\section{\gimsatul}
\label{sec:sharing-clauses}

\begin{figure}
\centering
\ifcompactfigures
\includegraphics[width=.9\textwidth]{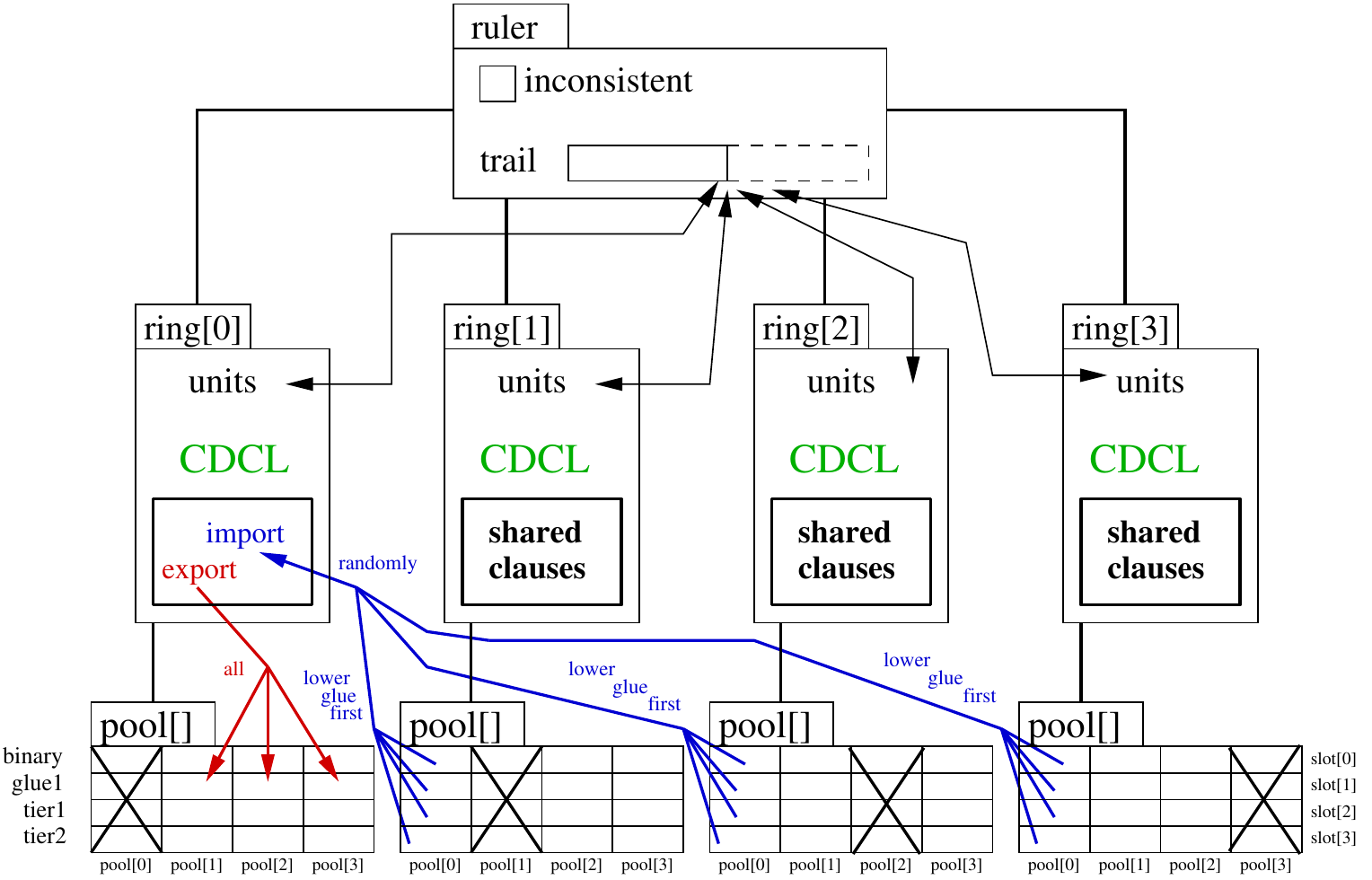}
\else
\includegraphics[width=\textwidth]{architecture}
\fi
\caption{\small Exporting and importing learned clauses, including the empty
clause through the ``inconsistency'' flag, learned units, and low glue
clauses.  Irredundant (including original) clauses are shared up-front
while cloning the initial clause data-base (after pre- and inprocessing).}
\label{fig:gimsatul-architecture}
\end{figure}

The key difference of our new solver \gimsatul compared to all recent portfolio solver
is the physical sharing of clauses.  In
order to do so, we have to revisit the implementation of watched literals, a data
structure to identify propagation and conflicts
(Section~\ref{sec:revis-prop}). Invariants on watched literals also limit how
clauses can be imported by an instance
(Section~\ref{sec:clause-shar-policy}).
We further  discuss how model
reconstruction is used to handle units (Section~\ref{sec:model-reconstruction})
and how sequential inprocessing can be achieved with clause sharing (Section~\ref{sec:inprocessing}).

\subsection{Overall Organization}
\label{sec:overall-organisation}

The core SAT solvers in \gimsatul follow
the design of our other recent solvers, particularly they
share many ideas with the ``sc2022-light'' version of \kissat.
\iftrue
For instance it includes
rephasing~\cite{rephase-pos20} and the stable and focused
mode.
\fi
There are two differences worth mentioning. First, the SAT
solver does not use an arena to compactly represent the clauses in memory.
One motivation for this trick is to put clauses consecutively in memory
in the order in which the solver accesses them~\cite{matesoos-arena}.
It further allows efficient garbage collection during clause data-base
reduction (forgetting heuristically-unimportant learned clauses).
This is something that we
cannot do for \gimsatul because each ring (solver instance) completely
independently allocates and reduces clauses in memory.
Hence, we allocate memory with \texttt{malloc}.  Second,
we use two watch lists, one for irredundant binary watch lists (shared across the
instances) and one for watching clauses, unlike the single watch list
used in our other solvers.

Overall the solving process looks as follows. First,
the main thread called \emph{Ruler} parses and allocates all clauses. At that point,
the \emph{Ruler} owns all the clauses. It then enters a ``cloning'' phase,
which first passes all its clauses to the
first ring as new owner. It also creates the shared global watch lists
for irredundant binary clauses. This first instance is then forked
into as many instances as needed, to match the number of requested solving
threads given on the command line, sharing all large clauses through new
sets of watched literal lists.
However, each solver instance reuses the same watch lists (a 
flat literal array) for the irredundant binary clauses.

After cloning, solvers start solving the CNF individually,
importing and exporting learned clauses (Section~\ref{sec:clause-shar-policy}).
The first solver instance to determine the problem as solved is
declared the winner and a
termination flag is raised, forcing other solver threads to exit
their CDCL loop too.
If the winner
deduced the empty clause the problem is unsatisfiable.  Otherwise it has
found a satisfying assignment
which is then extended by the main thread to a full model
using the reconstruction stack of the \emph{Ruler}/\emph{Simplifier} produced during pre- and inprocessing.

All solver instances logically reclaim (dereference) clauses satisfied by
learned root-level units, as part of frequent clause-database reductions,
which mainly have the purpose of discarding useless learned clauses.
However, reclaiming in this context just means decreasing the reference
count of a clause.  Only until the satisfying root-level unit has reached
all solver instances and is picked up during clause-database reduction,
that root-level clause is finally physically deallocated (and is also deleted
in the proof trace).

Cleaning the clauses from root-level falsified literals is much more
involved, as the actual clauses can not be shrunk in parallel -- the blocking and watched literals
potentially have to be changed.
For that purpose the first ring is responsible for starting simplification
rounds during which all irredundant clauses are handed over back to the
\emph{Ruler}.  Redundant watches pointing to redundant clauses are saved locally
for each ring separately.  After this preparation phase (called
  ``uncloning'' in the code), the inprocessing can start.

As clauses are shared we do not remove falsified literals during solving.
Instead solvers synchronize regularly and
\emph{hand} back their
clauses to the initial thread which then, in the role of the
\emph{Ruler} process, becomes responsible for ($i$) removing
satisfied clauses (which is something that can already be done by each ring
by marking the clause as removed locally), ($ii$) removing false literals from
clauses, and ($iii$) renumbering literals if holes appeared to keep literals
compact. After that, the heuristics are adapted (by renaming
literals in those heuristics too).

Besides sharing clauses and waiting for each other for inprocessing,
all solver instances are independent of each other and can run
using a different strategy. We currently use a simple and
limited portfolio: we use a different initialization for the
first random walk (limited local search) that initializes the phases,
which is one of the phases of our rephasing strategy~\cite{rephase-pos20}.
The result of this walk is exported and used as
saved phases initially, leading the solver to different search
directions. Interestingly, we initially had no diversification at all
and already observed improvement in the performance of the solver, due
the exchange of clauses (described in more details in Section~\ref{sec:clause-shar-policy}).
In Table~\ref{tab:portfolio-policy}, we show the amount of diversification we do, but
note, that we currently only have 12 different configurations.

\begin{table}
  \centering
  \begin{tabular}{r|cccc|cccc|cccc}
    Ring number mod 12&  0& 1& 2& 3& 4& 5& 6& 7& 8& 9& 10& 11\\
    \toprule
    Mode&S+F&S&F&S+F&S&F&S+F&S&F&S+F&S&F\\
    Phase&0&1&0&1&0&1&0&1&0&1&0&1\\
    Reason Bumping&0&0&1&1&0&0&1&1&0&0&1&1\\
  \end{tabular}
  \caption{\small Strategy of each thread (S = stable, F = Focused)}
  \label{tab:portfolio-policy}
\end{table}
\subsection{Revisiting Propagations}
\label{sec:revis-prop}

Since \zchaff~\cite{zchaff}, all modern SAT solvers use \emph{watched literals}
to identify clauses that can propagate or are in conflict. The idea is to
distinguish two literals in a clause. Whenever either of those literals is
set to false,
then the clause must be checked to see if it can propagate information or is
conflicting. Otherwise, the clauses do not need to be updated (nor visited).
This is essential to make propagations and conflict detection efficient.
Originally pointers to the two watched
literals in each clause were used. Modern
implementation however make sure that the first two literals of the clause are watched,
by swapping clause literals during propagation.

However, insisting on watching exactly the first two literals does not work when sharing clauses because
different solver threads (rings) have different partial assignments (and
trails) and thus usually watch different clauses and literals in those clauses.
Hence when sharing clauses we can not swap literals in the literal list of a
clause anymore.

Our solution is to store the watched literal pair
in a watcher data-structure separately, which also has a pointer to the immutable actual clause.
Those watchers are \emph{not} shared amongst solver instances, except for irredundant \emph{binary
clauses}, because they do not require any changes.
It is important to notice that, as the clauses are not changed, there
is no need for locks when accessing a clause by different threads.

The pair of literals in our thread-local watcher data-structure also serves as
``blocking literals'' which are checked first to satisfy the clause.
This is
a common technique to reduce the number of times the actual clause
data has to be accessed.  If the blocking literal check fails, however,
this scheme incurs an additional pointer access compared to the standard
version of placing the two watched literals in front of the literal list of
a clause, effectively merging the clause and the watcher data-structure.
On the other hand our watcher structure is much more compact than a full
clause and thus likely has better cache locality.

\subsection{Clause Sharing}
\label{sec:clause-shar-policy}

An important design choice for
parallel solvers is to determine which clauses to share and
how often they are imported. We use a very simple policy: ($i$) at most
one clause is imported before each decision and ($ii$) when learning a clause
with a low glucose level, we immediately export it.
We do however import and export all derived root-level assignments eagerly
as well as check for termination and thus ``inconsistency'' (another thread
proved the formula to be unsatisfiable).
Figure~\ref{fig:gimsatul-architecture} shows how the sharing happens and is described in the text below.
All rings share the trail composed of unit literals.

\PARAGRAPH{Which Clauses to Import.}
When importing clauses before a decision, only a single one is imported at a time.
To import and export clauses, each ring has 4 slots for each
other ring with clauses to export: one for binary clauses (64 bits representing
two 31-bit literals), glue 1 (non-binary) clauses, glue-2 clauses (remaining
tier1 clauses), and
finally tier2 clauses (glue 3 to 6). In
Figure~\ref{fig:gimsatul-architecture}, the slots are called
\emph{pools} and the pool from a ring for itself is crossed out (because
it is useless to have a ring share a clause with itself).

Before every decision, each ring attempts to import one clause after
checking that no new units should be imported first.
Without any new unit it selects its pool among the shared pools of another randomly chosen ring.
The goal of randomly picking the exporting ring is to implement
a global (bounded) queue with relaxed semantics~\cite{kstack}, i.e., a
bounded $k$-queue, which probabilistically guarantees low contention.
Each queue is implicitly bounded because threads during exporting learned clauses simply overwrite
references and thus drop clauses with the same glue class (binary, tier1, tier2, tier3)
as the exported clause.

Then the importing ring
checks its slots in order with lowest glue first (this is a fast-path
without locking) if there is any clause to import (the blue arrows in Figure~\ref{fig:gimsatul-architecture}).
If one is found, then the slot is emptied
with an atomic exchange operation (thus requiring 64-bit word size).

\PARAGRAPH{Importing Clauses.} As described in the previous paragraph, the
solver imports one clause at a time.
\iffalse
There are several cases:

\begin{itemize}
  \item the clause neither propagates nor is conflicting. In that case the
        watched literals must be selected according to the trail (either a true
        literal and an unset literal, two unset literals, or a true and a false
        literal such that the level of the true literal is smaller or equal to
        the level of the false literal).
  \item the clause is conflicting. In that case we backtrack to the level where
        it is a conflict (i.e., to the maximum level of the level of all
        literals in the clause).
  \item the clause is propagating. In that case we backtrack to the point where
        the propagation should have take place and continue from there on. The
        backtracking could be avoid if we would instead use chronological
        backtracking~\cite{chrono1, chrono2}, but we leave this as future work.
\end{itemize}
\else
The current partial assignment on the trail has to be fixed if the clauses is
propagating/conflicting. Otherwise, some literals (not arbitrary ones) have to
be selected as watched literals to fulfill the watch list invariants.
\fi

Before we actually import a clause it is checked for not being already
subsumed by another existing clauses, using the watch list as an approximation
for occurrence lists.  This forward subsumption check
traverses the watcher list of a new watched literal which is smaller
and thus might miss some subsuming clauses, but is complete for exact
matches (identical clauses ignoring root-level falsified literals).
Besides adapting the current interpretation to the single imported clause, importing simply means
watching the correct literals and adding the clauses to the correct watch lists.

In our implementation importing a clause amounts to increasing the overall
number of occurrences of that clause. We rely on atomic operations to
adjust these counters (using \texttt{atomic\_fetch\_add} and
\texttt{atomic\_fetch\_sub} from \texttt{stdatomic.h} in C11).

In contrast to operating in single-threaded mode
and unlike all our single-threaded SAT solvers \gimsatul using multiple
threads is highly
\emph{non-deterministic}, because importing clauses is done eagerly and depends
on the exact thread scheduling, in which order memory accesses occur and
caches are updated etc.

\PARAGRAPH{Exporting Clauses.}
Important learned clauses with small glucose level are exported immediately during
the conflict analysis, with
references to the exported clause added to the $n-1$ pools of
the exporting thread (the red arrows starting from ring 0 in Figure~\ref{fig:gimsatul-architecture}).
Each pool corresponds to exactly one thread and has four slots 
of clauses references sorted by glucose level (binary,
tier1, tier2, tier3 clauses).
This allows to share these important learned clauses with all other
solver threads, prioritized by importing low-glue clauses first.

\PARAGRAPH{Life and Death of Clauses.}
Remark that the clauses are imported according to the score they had previously
and are handled the same way as every other learned clause. Therefore, low LBD
clauses are never removed (in particular binary clauses), but unused tier2
clauses will eventually be deleted.  Large tier1 clauses might be removed
by vivification though.

\iftrue
During execution, the LBD score of clauses can actually change. This update is
for example in \glucose for clauses involved in the conflict analysis (and only
if the score decreases), although the actual LBD of the score is already defined
when the clause is propagating. An interesting question is whether
\emph{promoted} clauses, i.e., clauses whose LBD changes enough to become tier1
or tier2 clauses should be shared. We experimented with promotion, but did not observe
any benefit.  It is also rather difficult to implement as it might either
result in glue values to diverge between clauses and their watches or otherwise
requires atomic update of glue values in clauses.
\fi

\subsection{Model Reconstruction}
\label{sec:model-reconstruction}

\gimsatul relies on the model reconstruction~\cite{DBLP:conf/sat/JarvisaloB10} to
produce a model and ``undo'' the inprocessing. This reconstruction
stack is not shared amongst the rings. Instead, a single
one is used. This is sufficient because all solvers work on the same
formula.

Unlike previous solvers, we actually go one step further and
completely remove fixed literals assigned at root-level (decision-level zero).
In \kissat we would also
remove those literals but not remove them from external partial assignment
Therefore, we do not need to put those
literals on the reconstruction stack. Reusing the reconstruction stack
is in particular used for units derived during inprocessing
techniques (see more details in the next paragraph), because it avoids
communicating such literals back to each SAT instance.

\subsection{Inprocessing}
\label{sec:inprocessing}

We have implemented two different kinds of inprocessing in \gimsatul. Some
transformations are part of probing like vivification. They run directly in the
different rings but they do not shorten shared clauses. Instead a new
shortened clause is added and the other is removed from the clause set. Other
inprocessing techniques require to change the set of original or in general
irredundant clauses such as
bounded variable elimination~\cite{DBLP:conf/sat/EenB05}. For those
transformation, each instance first gives up all references to the clauses. At
the end, only the first instance knows the location in memory of all the
clauses. It gives them back to the \emph{Ruler} instance which then starts
the \emph{Simplifier}.
%
This \emph{Simplifier} is then in charge of transforming all the
irredundant clauses (including shortening and strengthening them) using the
standard algorithms, e.g., for variable elimination. After that the clause are
passed back to the first ring which in turn passes them back to the
other rings (with the shared watch list of binary clauses).

Getting this ``uncloning '' option to work was rather challenging, because
units produced must still be shared amongst all rings. This in turn
can enable more propagations at root-level (at decision-level zero) that again needs to be shared.
If this is not done properly, some units
might get lost, which is an issue as then the candidate model of the different
instances might not know about those units, leading potentially to incorrect
models: If the \emph{Simplifier} removes irredundant clauses containing a
given literal, because this literal is not assigned and can appear in the redundant clauses, it can get assigned to the
opposite value leading to an incorrect model.

\section{Experiments on Scalability}
\label{sec:scale-experiments}

We ran experiments\footnote{Experimental data 
available at \url{https://cca.informatik.uni-freiburg.de/pos22gimsatul}.} on our cluster equipped with Intel Xeon
E5-2620 v4 CPU at \SI{2.10}{\giga\hertz} (with turbo-mode disabled)
with a memory limit of \SI{128}{\giga\byte} for each node.  Those CPUs
have 16 real cores and 32 cores using hyper-threading.  In order
to keep the testing time reasonable (and running solvers in parallel), we assigned
\SI{127}{\giga\byte} for 8 cores or more, \SI{63}{\giga\byte} for 4
cores, \SI{31}{\giga\byte} for 2 cores, and \SI{15}{\giga\byte} for 1
core.

\begin{table}
  \centering
  \begin{tabular}{r|ccc|S[table-format=7.0]S[table-format=7.0]S[table-format=9.0]}
    & solved & sat & uns & {elapsed time} & {PAR-2} & {space}\\
    & &  &  &{(s)}  & {(\(10^{3}\))} & \\  \toprule
        \gimsatul-32 & 310& {\bf 151}& 159& 3928732& 1031& 1578816\\
\pmcomsps-32 & {\bf 315}& 141& {\bf 174}& 5804244& 1037& 3038965\\
        \gimsatul-16 & 308& 149& 159& 2069523& 1055&  843444\\
\pmcomsps-16 & 309& 136& 173& 2607129& 1076& 1495201\\
        \gimsatul-64 & 304& 150& 154& 4768809& 1119& 2948359\\
         \gimsatul-8 & 298& 144& 154& 1223199& 1178&  460013\\
 \pmcomsps-8 & 297& 131& 166& 1584009& 1229&  723740\\
         \gimsatul-4 & 282& 138& 144&  744566& 1371&  263227\\
 \pmcomsps-4 & 260& 119& 141&  854403& 1614&  317691\\
         \gimsatul-2 & 262& 130& 132&  499420& 1633&  167771\\
         \gimsatul-1 & 230& 117& 113&  263851& 1963&   80947\\
    \pmcomsps-2 & 204&  87& 117&  679142& 2187&  187676\\
    \bottomrule
  \end{tabular}
  \caption{\small Results on the problems from the SAT Competition 2021}
  \label{tab:sc2021-results}
\end{table}

\ifcompactfigures
\else
\begin{figure*}
  \centering
  \includegraphics[scale=0.6]{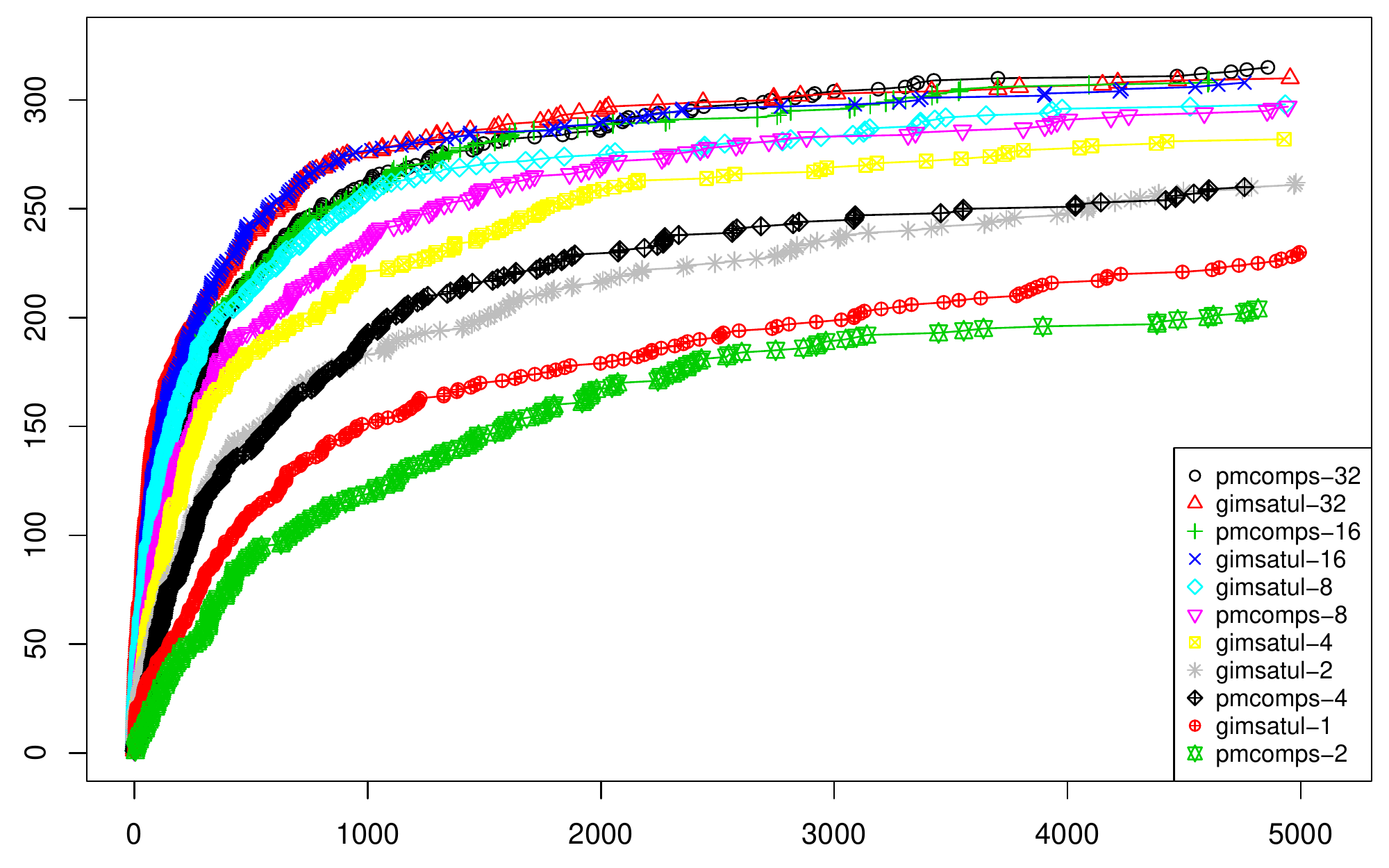}
  \caption{\small Cumulative distribution function (CDF) of the wall-clock solving
  time of both \gimsatul and \pmcomsps for various considered number of threads.}
  \label{fig:cdf-gimsatul-pmcomsps}
\end{figure*}
\fi

\PARAGRAPH{Performance.}
We first compare the overall wall-clock-time performance of our solver
\gimsatul to the
state-of-the-art solver \tools{p-mcomsps}~\cite{pmcomsps}, which won the 
parallel track of the SAT Competition 2021. It features an advanced SAT
core solver and incorporates many ideas to improve parallel solving including
sophisticated diversification techniques. However, it fails when run with a single
thread (producing an exception). We assume that this is due to the fact that one thread is used to strengthen clauses
and the others for solving.  Therefore we only consider experiments for
\tools{p-mcomsps} with at least two threads.
The CDF (Figure~\ref{fig:cdf-gimsatul-pmcomsps}) and the raw results (Table~\ref{tab:sc2021-results})
give the results for this initial experiment.

At first we were surprised by the fact that \pmcomsps needs twice as many
threads to match the performance of \gimsatul for less than 8 threads.
Due to time constraints, we could not run all the configurations on the SAT
Competition 2020 too.  But partial runs show that the results are
very similar for 2, 4, and 8 threads (i.e., the performance of
\pmcomsps with \(n\) threads is similar to the performance of
\gimsatul with \(\nicefrac{n}{2}\) threads).
This shows that our results do not
seem to be biased to the 2021 benchmarks
(a valid threat to validity).

The plot also shows that
\gimsatul is in general faster than \pmcomsps,
particularly with respect to the PAR-2 score,
even though both solve a similar number of benchmarks.

We have also experimented with higher number of threads than our machines
support (Figure~\ref{fig:cdf-gimsatul-saturated}). The performance gap when
using all 32 virtual threads on the 16 ``real'' cores does not yield a
performance decrease. However, when using 64 threads, performance decreases.
This indicates that \gimsatul does not spend all its time waiting for the other
threads.

\ifcompactfigures

\begin{figure}
\centering
\begin{minipage}[t]{.48\textwidth}
  \includegraphics[width=\linewidth]{experiments/sc2021/analyze/cdf-gimsatul-pmcomps-on-sc2021-benchmarks.pdf}
  \caption{\small Cumulative distribution function (CDF) of the wall-clock solving
  time of both \gimsatul and \pmcomsps for various considered number of threads.}
  \label{fig:cdf-gimsatul-pmcomsps}
\end{minipage}\hfill
\begin{minipage}[t]{.48\textwidth}
  \centering
  \includegraphics[width=\linewidth]{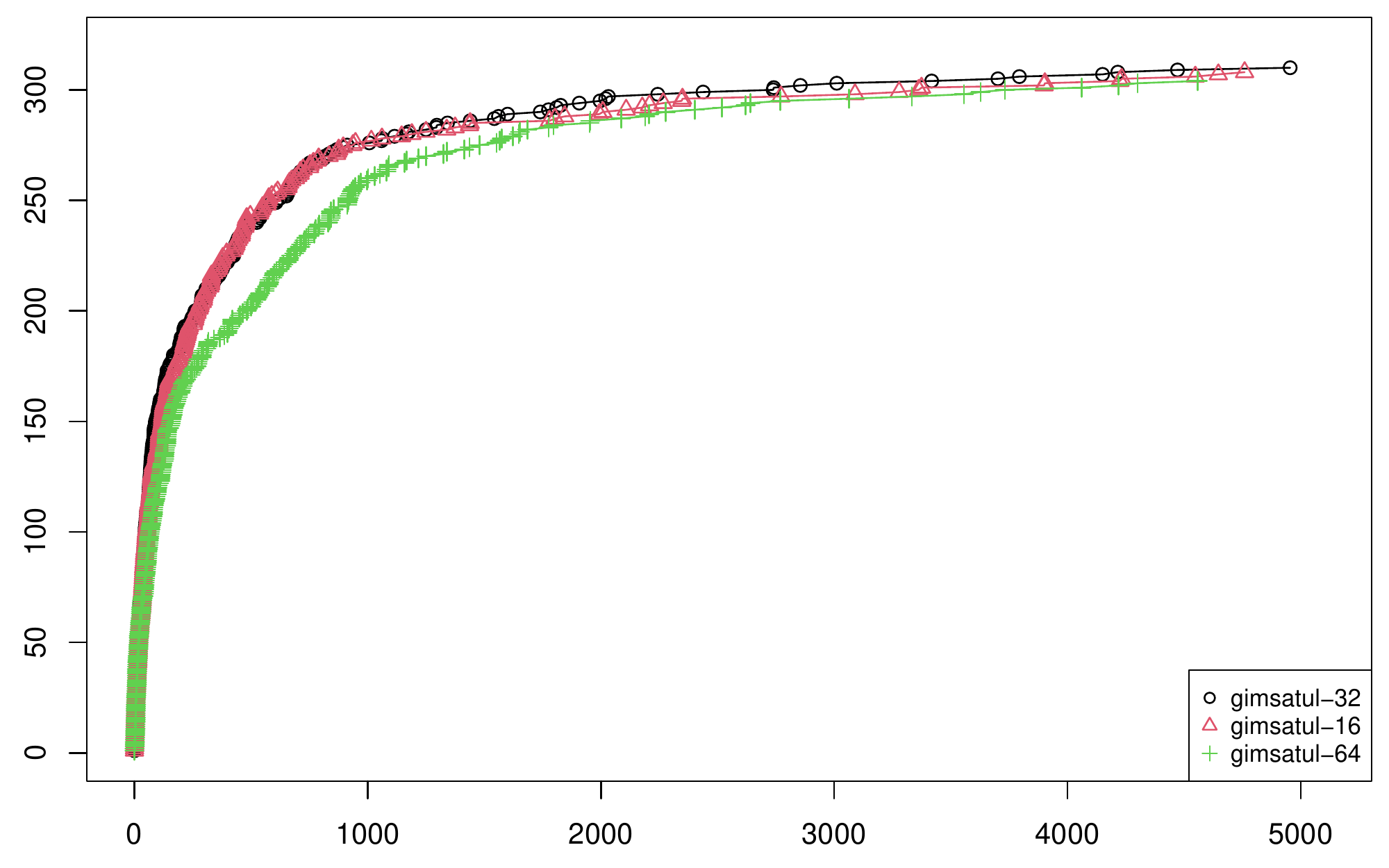}
  \captionof{figure}{\small CDF with more threads than cores}
  \label{fig:cdf-gimsatul-saturated}
\end{minipage}
\end{figure}

\begin{figure}
\centering
\begin{minipage}{.48\textwidth}
  \centering
  \includegraphics[width=\linewidth]{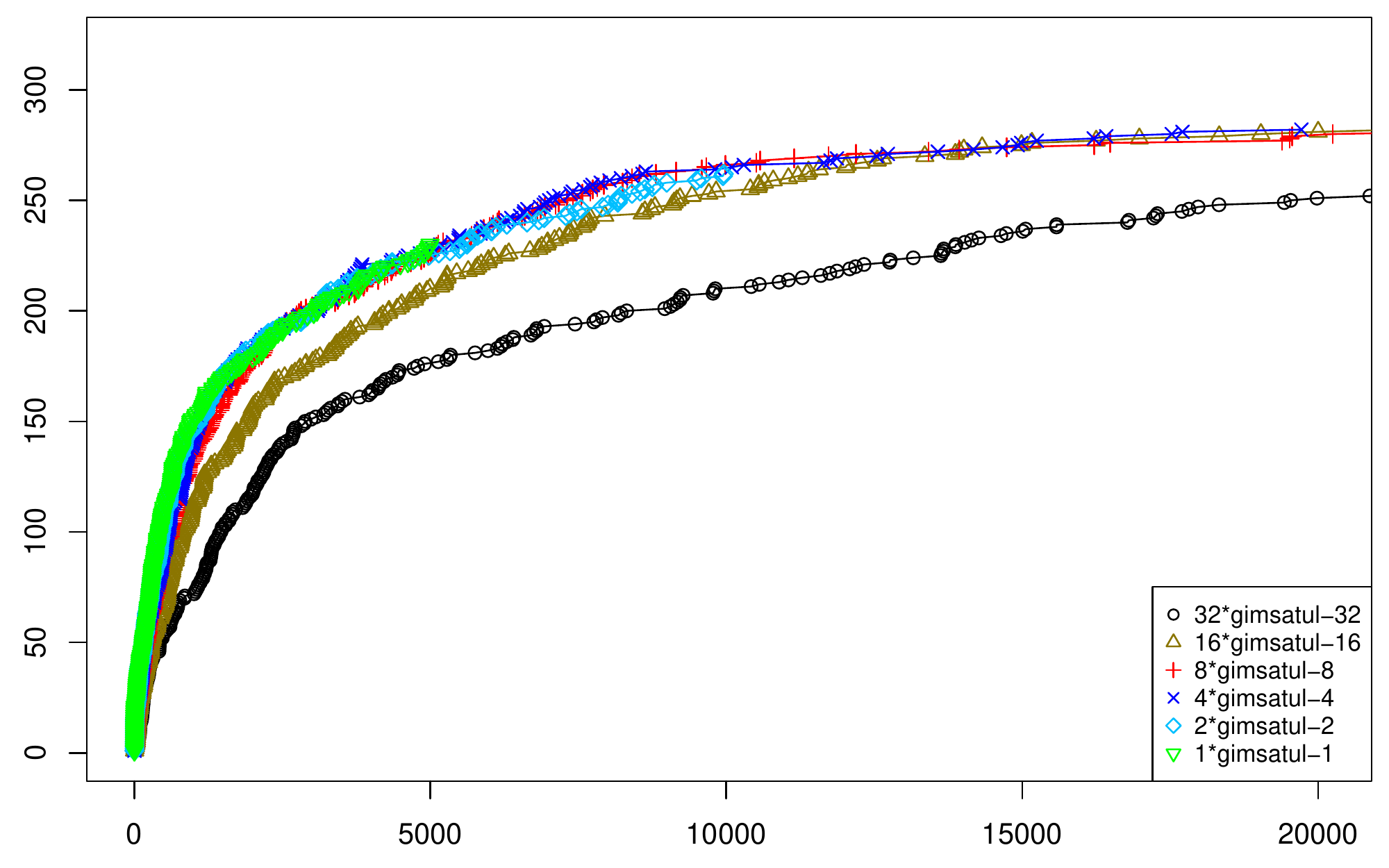}
  \captionof{figure}{\small Scalability \gimsatul}
  \label{fig:cdf-gimsatul-scalability}
\end{minipage}\hfill
\begin{minipage}{.48\textwidth}
  \centering
  \includegraphics[width=\linewidth]{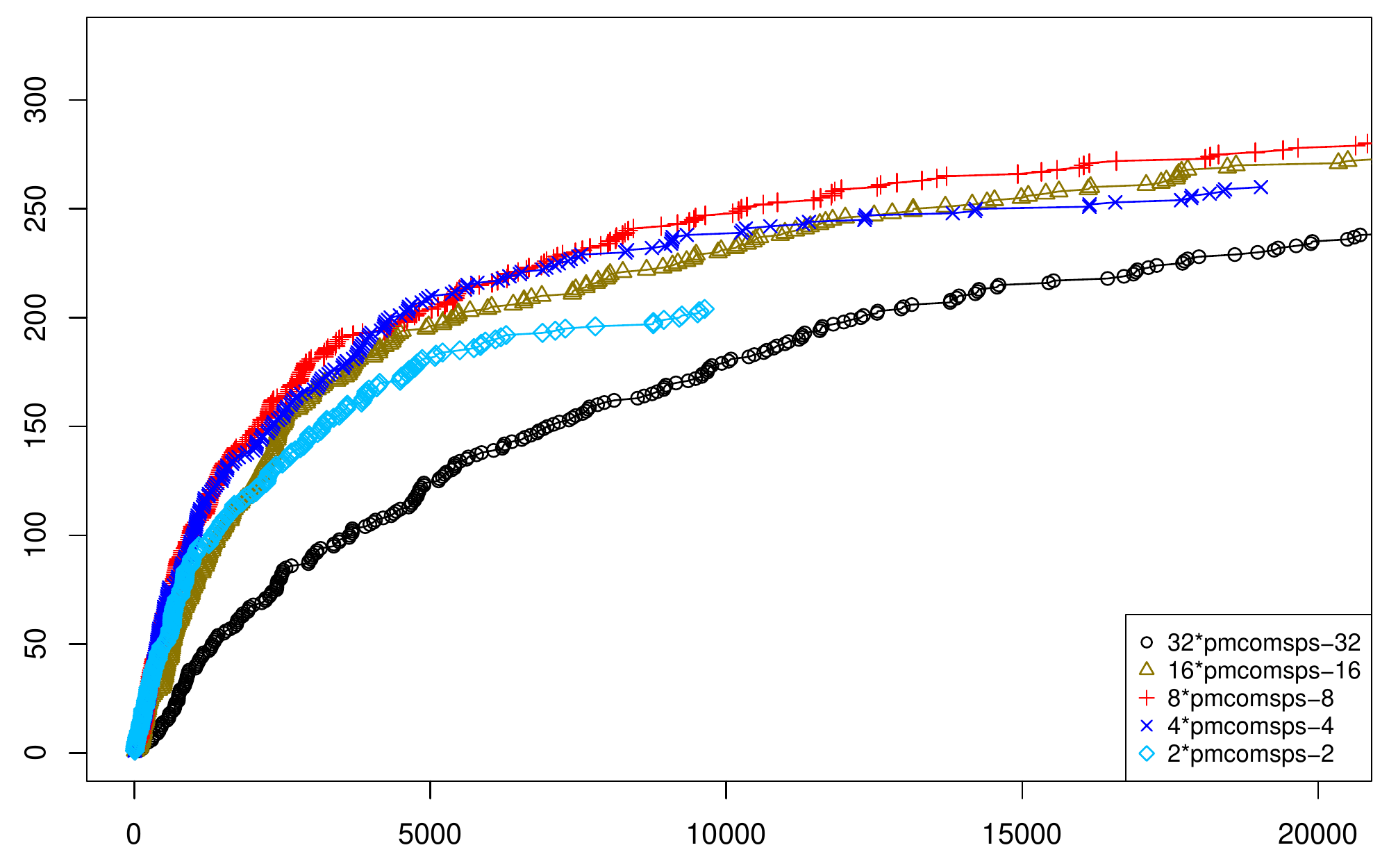}
  \captionof{figure}{\small Scalability \pmcomsps}
  \label{fig:cdf-pcomsps-scalability}
\end{minipage}
\end{figure}

\begin{figure}
\centering
\begin{minipage}{.48\textwidth}
  \includegraphics[width=\linewidth]{experiments/sc2021/analyze/cdf-gimsatul-pmcomps-on-sc2021-benchmarks.pdf}
  \captionof{figure}{\small Cumulative distribution function (CDF) of wall-clock solving
  time of \gimsatul and \pmcomsps for various number of threads.}
  \label{fig:cdf-gimsatul-pmcomsps}
\end{minipage}\hfill
\begin{minipage}{.48\textwidth}
  \includegraphics[width=\linewidth]{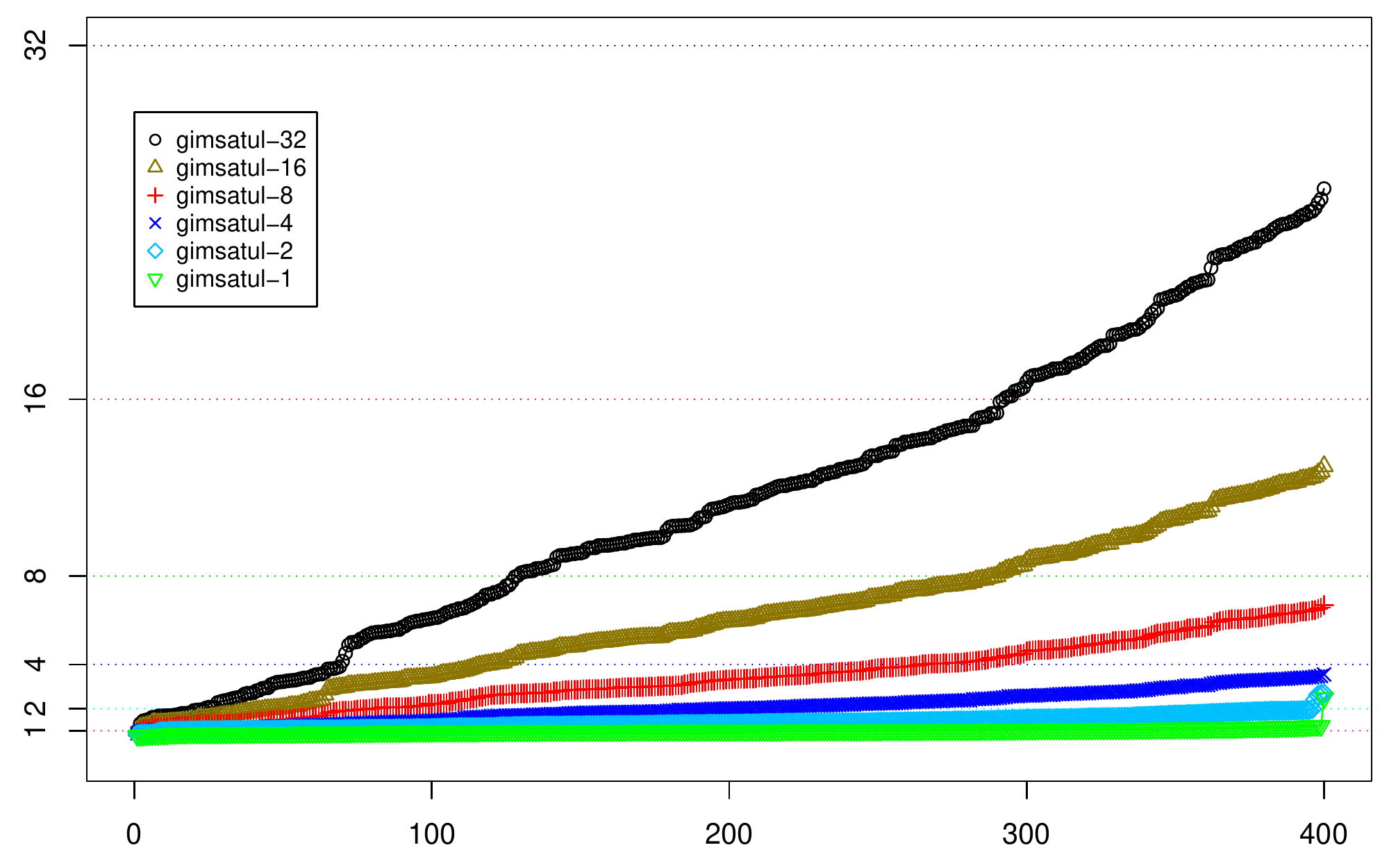}
  \captionof{figure}{\small Initial relative memory increase of \gimsatul
  w.r.t.~resident
  size before and after cloning, i.e., after preprocessing but
  before solving.}
  \label{fig:mem-usage-initial-increase}
\end{minipage}
\end{figure}

\else

\begin{figure*}
  \centering
  \includegraphics[scale=0.6]{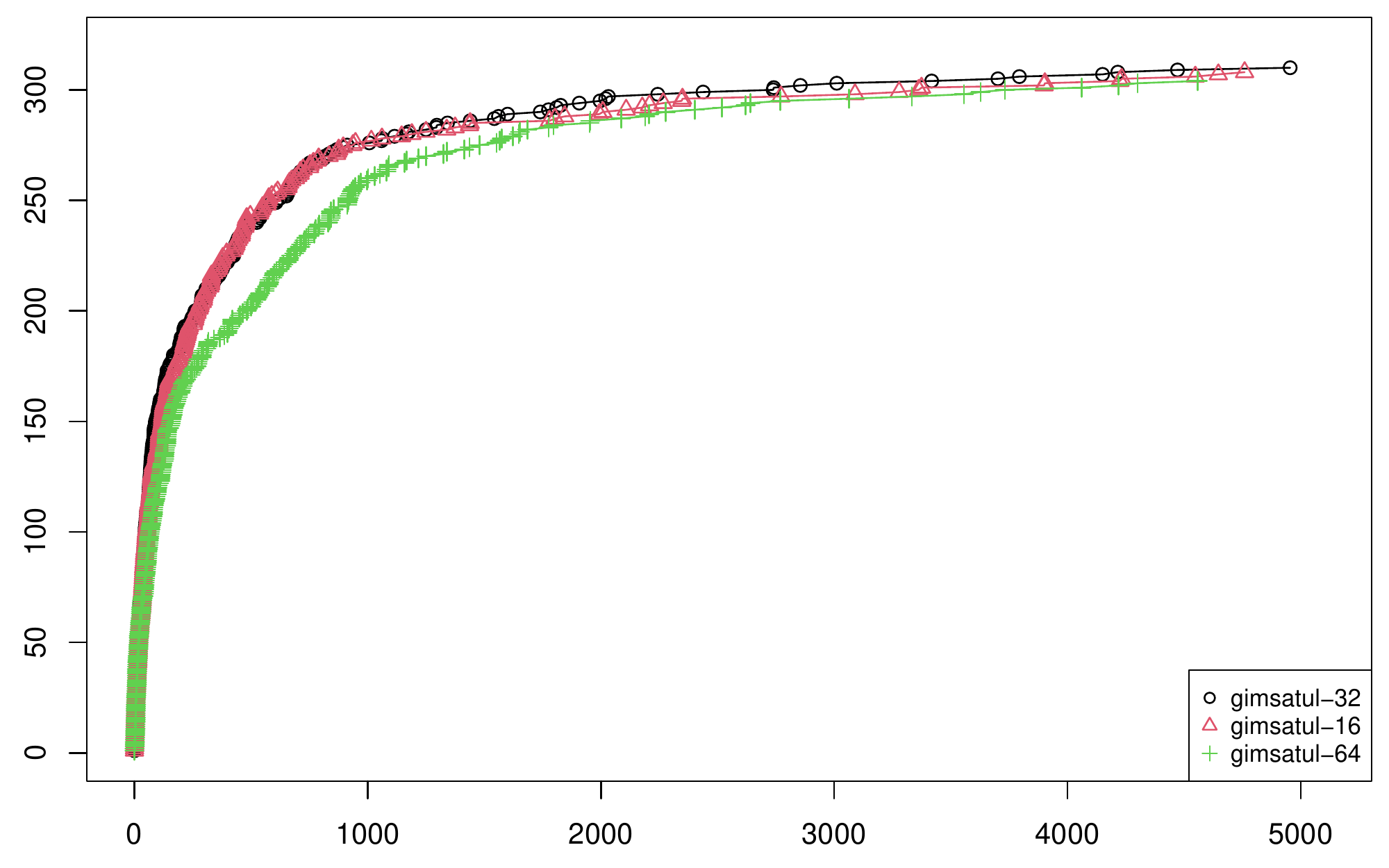}
  \caption{\small CDF with more threads than cores}
  \label{fig:cdf-gimsatul-saturated}
\end{figure*}

\begin{figure*}
  \centering
  \includegraphics[scale=0.6]{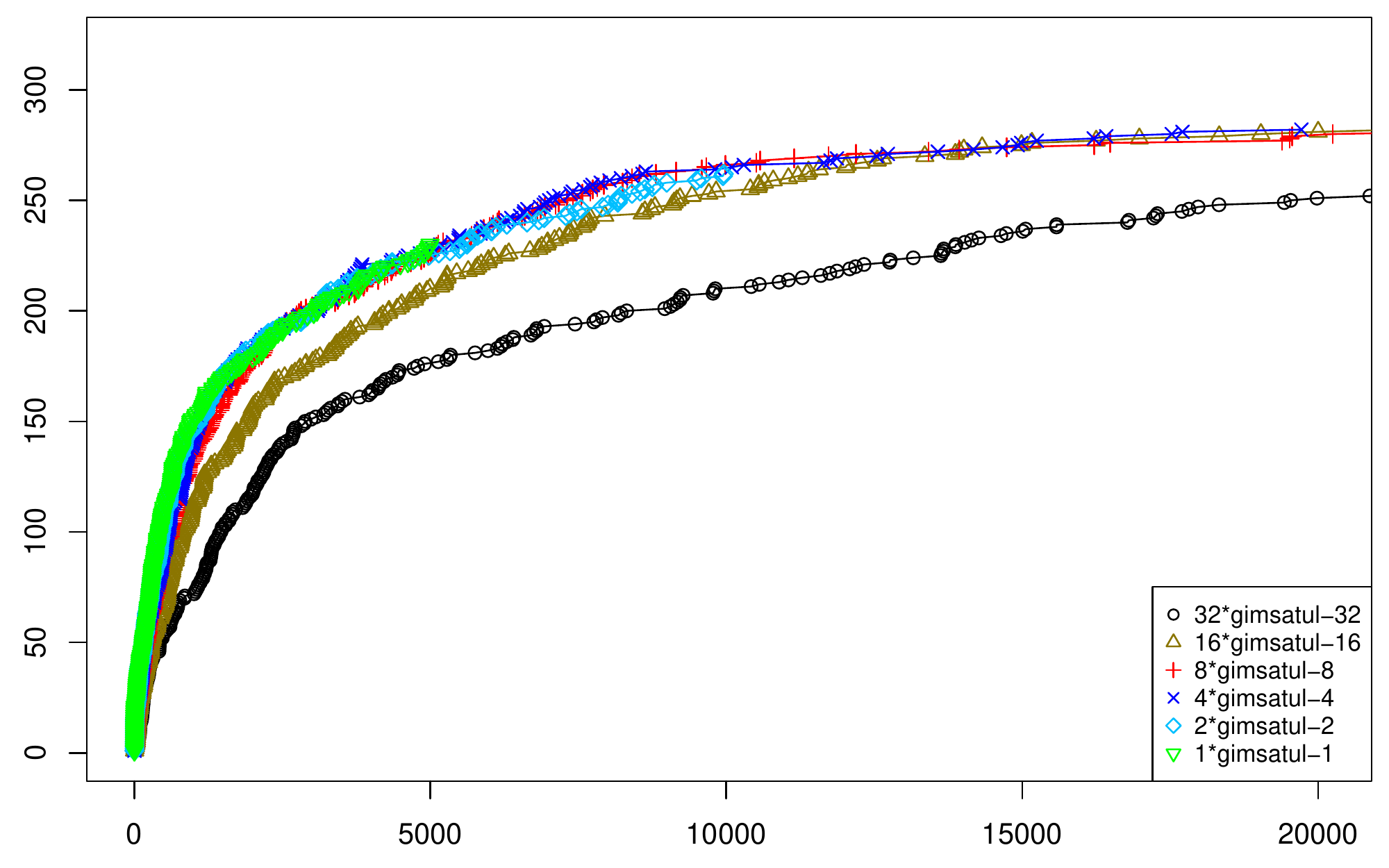}
  \caption{\small Scalability \gimsatul}
  \label{fig:cdf-gimsatul-scalability}
\end{figure*}

\begin{figure*}
  \centering
  \includegraphics[scale=0.6]{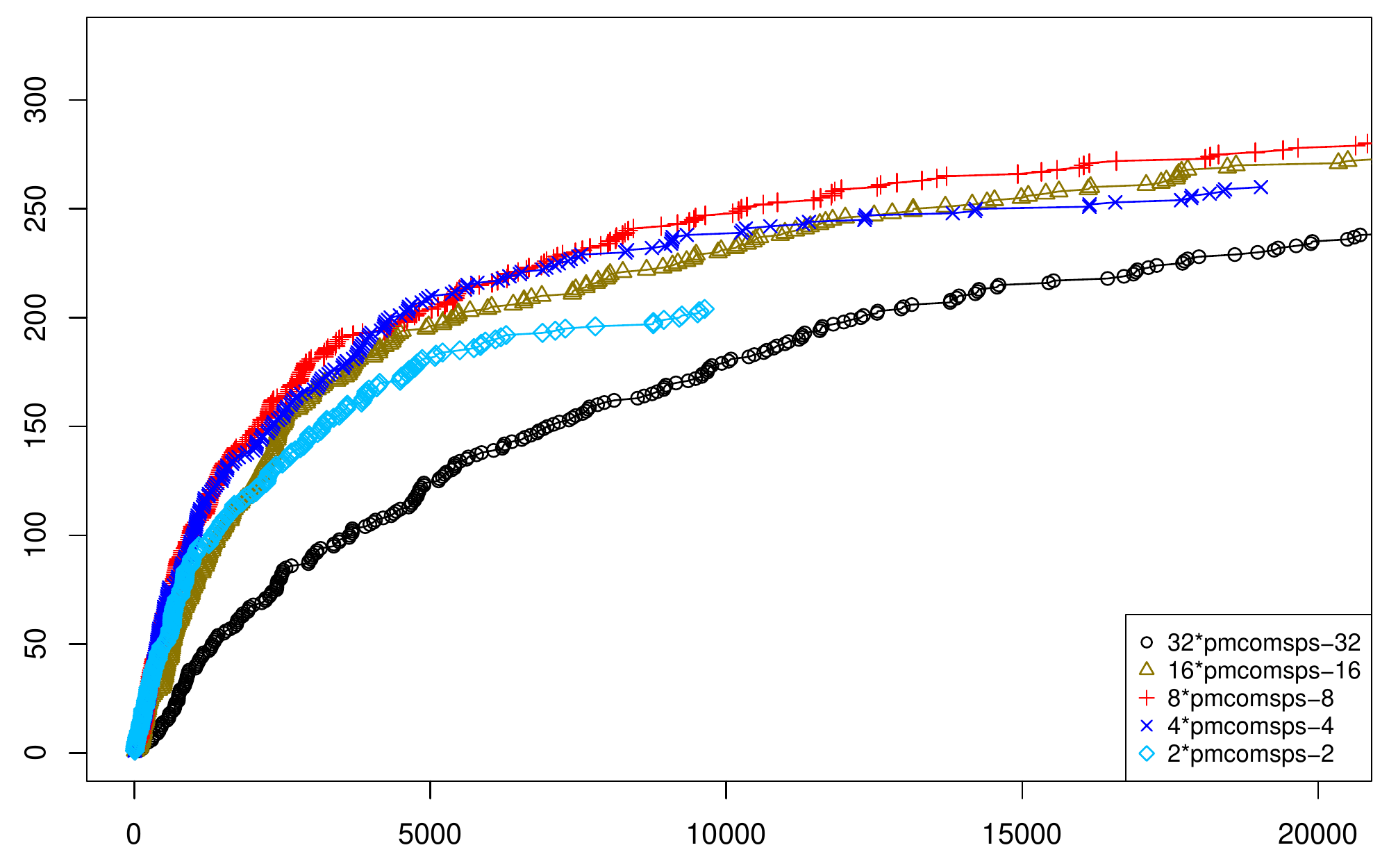}
  \captionof{figure}{\small Scalability \pmcomsps}
  \label{fig:cdf-pcomsps-scalability}
\end{figure*}

\fi

\PARAGRAPH{Scalability.}
While we only considered wall-clock-time performance above it is also interesting
to investigate how effectively compute resources are used.
This kind of question can arise in a cloud
context where customers pay only for what is used and do not want to
provision redundant compute resources (cores) unless solving latency
is reduced effectively.
We try to answer this question by
plotting CDFs
where (wall-clock) time is replaced by ``\(\text{time} * \text{number of cores}\)''.
This is equivalent to run our multi-threaded version on a single core.

For \gimsatul (Figure~\ref{fig:cdf-gimsatul-scalability}), we see once
saturation is reached (i.e., after 32 threads on our 16 ``real'' core
nodes), performance decreases. The 16-thread version has a disadvantage at the
beginning, but catches up. One reason might be that our portfolio has only 12
different policies, and the 4 other rings run the same first four policies
again. Even though the behavior is not deterministic, more diversification is
probably better. It is significant to see that the other curves are basically
identical. This shows that our solver scales with the number of cores.
\iftrue
As future work we want to repeat this experiment with disabled
diversification (portfolio).
\fi

For \pmcomsps (Figure~\ref{fig:cdf-pcomsps-scalability}), the picture
is different. Remark that the 2-thread version stops at \SI{11000}{\second}, i.e., exactly the timeout of \SI{5500}{\second} when you have 2 threads.
First the performance loss for 32 threads is already much more
pronounced. So virtual threading seems to be very harmful. Second, we can see
that performance for 2 threads is worse than for the other configuration. We
attribute this to a lack of optimization and diversification for this case.
Third, peak performance seems to be reached already for 8 threads, but the
performance decrease compared to 16 threads is limited: it is more important to
go from 4 to 8 threads, than to go from 8 to 16.

\begin{figure*}
  \centering
    \begin{subfigure}[b]{0.45\textwidth}
      \includegraphics[scale=0.6]{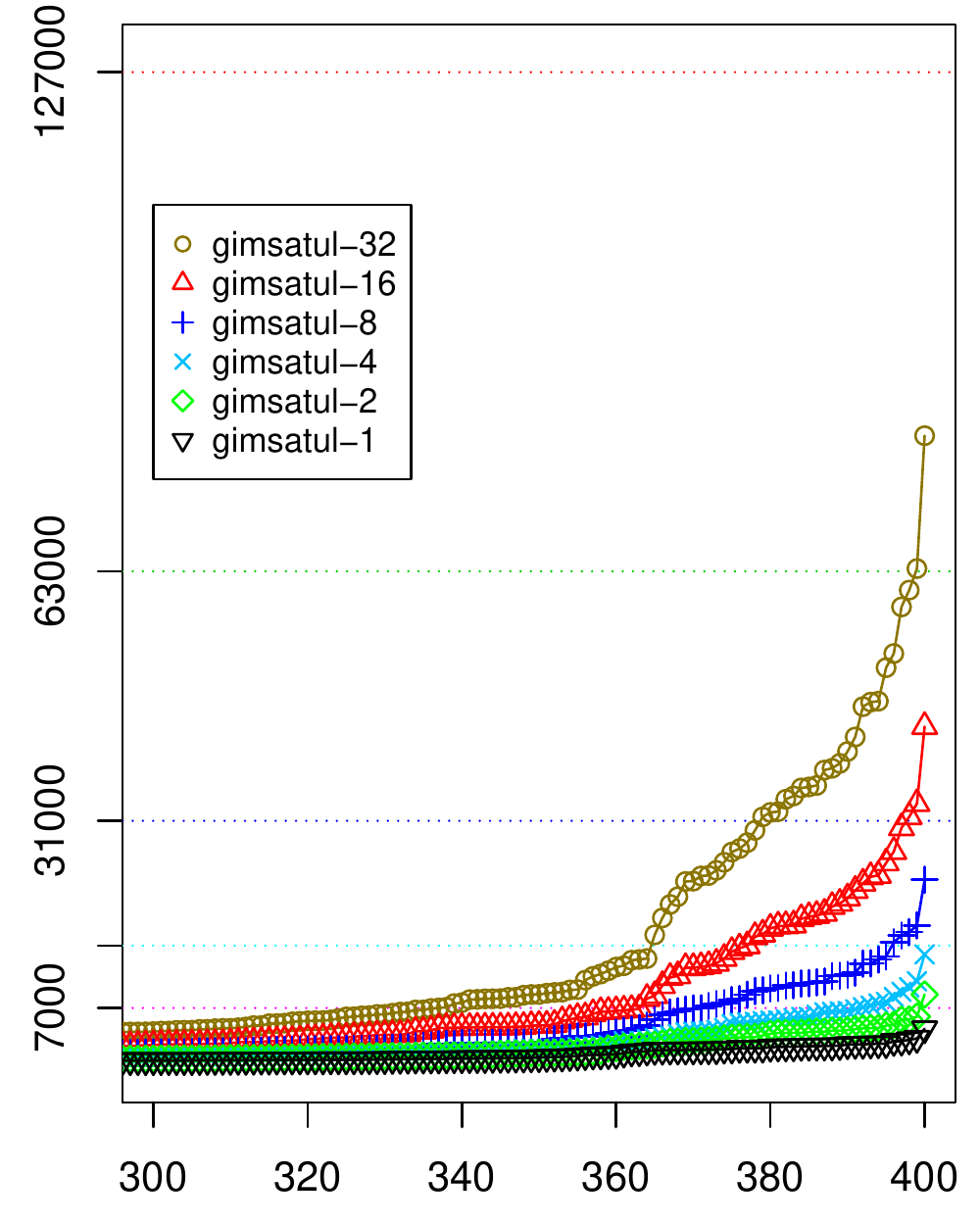}
        \caption{\small Memory usage of \gimsatul}
        \label{fig:mem-usage-gimsatul}
      \end{subfigure}
      \quad
    \begin{subfigure}[b]{0.45\textwidth}
      \includegraphics[scale=0.6]{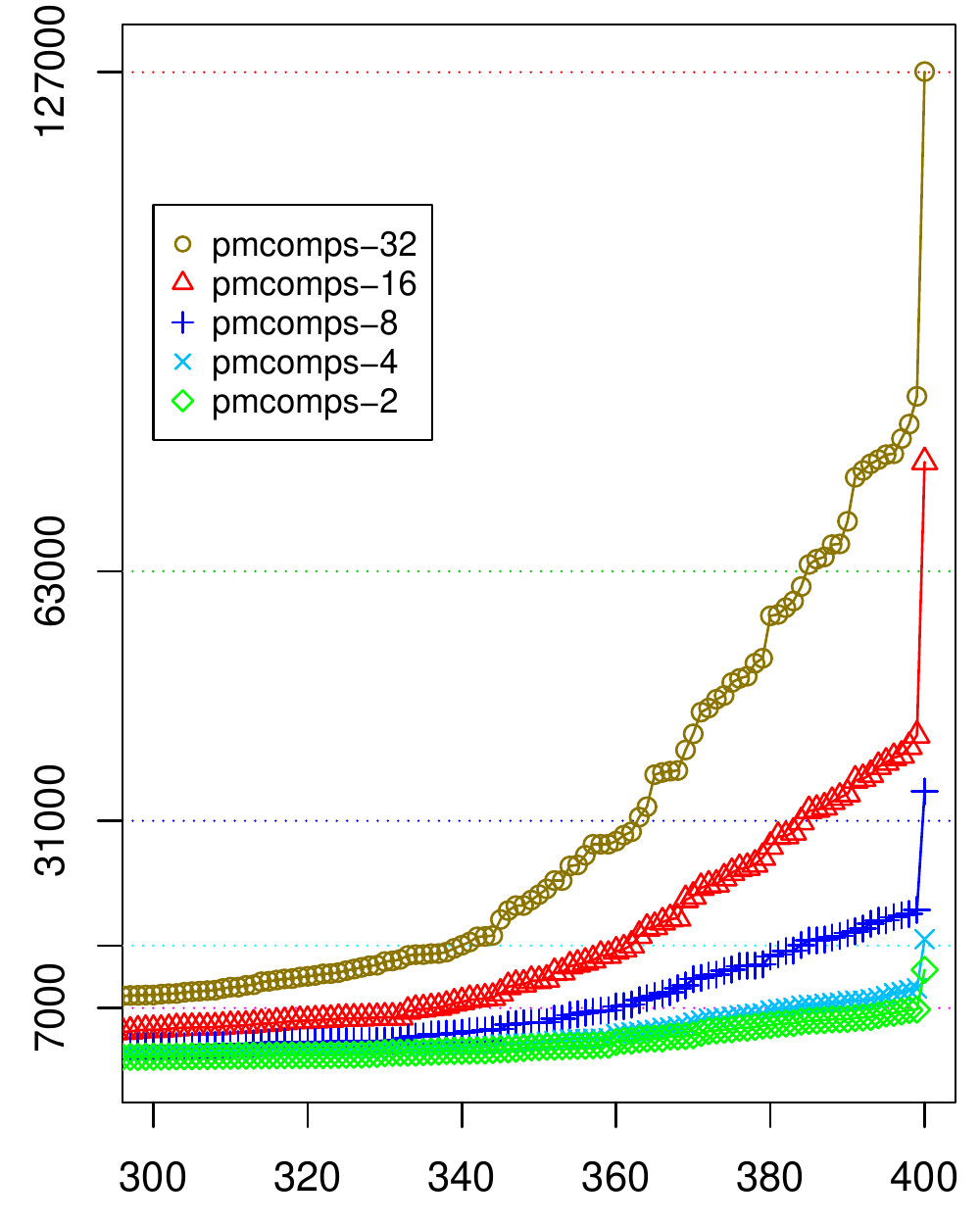}
        \caption{\small Memory usage of \pmcomsps}
        \label{fig:mem-usage-pmcomsps}
      \end{subfigure}
      \caption{\small Memory usage in MB of both solvers on problems of the SAT Competition 2021}
      \label{fig:mem-usage}
\end{figure*}

\PARAGRAPH{Memory Usage.}
In order to compare memory usage we have prepared two different plots. The first
(Figure~\ref{fig:mem-usage}) shows peak memory usage (maximum
resident-set-size) during the run. We can clearly see that \gimsatul uses
much less memory than \pmcomsps.  For 32 threads \gimsatul only needs in one
case for 32 threads slightly more than 80 GB and otherwise stays below 64
GB, which is half the memory available on our cluster nodes.
On the other hand,
also for 32 threads,
\pmcomsps hits the memory limit of 128 GB once.

In order to check how sharing works, we also checked the amount of memory used
after the first preprocessing round but before cloning and any learning is done
(Figure~\ref{fig:mem-usage-initial-increase}). This value is interesting because
it shows an (optimistic) view on the cost of duplicating the watch lists and all other
data-structures of the solver. With only binary clauses, the overhead would be very low.
Without any binary clause all watch lists are duplicated, even though the
actual clause data - the literals in clauses - are shared.
Here we show how much more memory is used once
\gimsatul has initialized all the different rings. We can see that the increase
is on average much lower than the number of solver instances, i.e.,
the number of threads.

\ifcompactfigures
\else
\begin{figure*}
  \centering
  \includegraphics[scale=0.6]{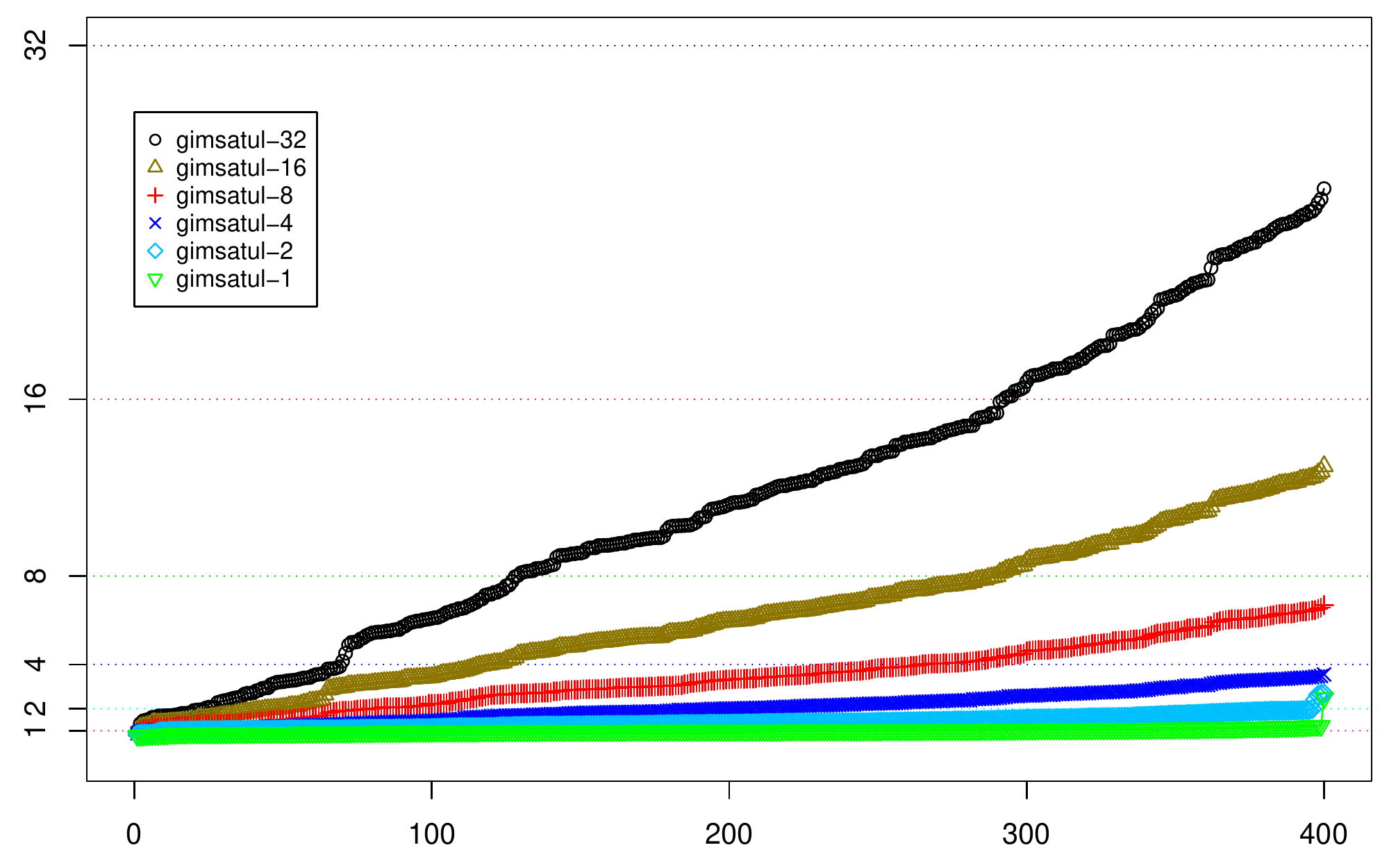}
  \caption{\small Initial relative memory increase of \gimsatul in terms of resident
  size before cloning and after cloning, i.e., after preprocessing but
  before any solving/learning takes place}
  \label{fig:mem-usage-initial-increase}
\end{figure*}
\fi

\section{Generating DRAT Proofs}
\label{sec:drat-proof-checking}

In essence, a DRAT proof certificate is a list of derived clauses ending with the
empty clause (if the problem is deemed unsatisfiable). The key idea is that
adding
new derived clauses is satisfiability preserving: if the initial problem has a model, then the
model with the new clauses has a model. Such clauses are said to be
\emph{redundant}.
In general such certificates have a multiset semantics: the checker keeps the clause
as many time as it was added (and removing one of the copies only removes one of them, not all).
In \gimsatul, we know how often a clause is present (so we can keep only one clause) but this
is not possible in general for parallel SAT solvers, requiring a copy for each shared clause.

The DRAT proof format was selected to be easy to implement for single threaded
SAT solvers: It is sufficient to dump the derived clauses in order and to stop
when the empty clause is derived.
For parallel SAT solvers, it is tempting to produce one proof file per solver
instance, but this does not work when the different instances exchange
clauses.
Instead we focus on a variant that produces a \emph{single} proof file. This
approach is not a new approach, but we did not find a description of it in
related work.

One trivial solution is that every derived clause is written to the proof file
in the order it is created and additionally added again (with multi-set
semantics) when the clause is imported by another solver thread.
This makes sure that the individual solver instances can delete
individual clauses as they like and log these deletion steps
independently of each other in the proof file.

With this trivial ``copying'' solution clauses are repeated as many times as they
are copied, increasing the size of the proof file almost linearly with
respect to the number of threads (core solver instances).
We propose instead in this work to
\emph{share} those clauses in the proof instead of duplicating them,
which in turn requires to share them among the solver threads too.
Unfortunately, this requires some substantial changes to the
data-structures for watched literals, as described in the next
section, and is one of the main reasons we started \gimsatul from scratch
instead of incorporating similar ideas into our sequential state-of-the-art
solver \kissat.

Generating proofs is very easy in \gimsatul: Every large (non-binary)
clause has an atomically incremented and decremented reference counter for
the number of occurrences in different solver threads (rings), occurrences
in clause pools and during simplification phases in the simplifier.  When creating
(allocating) the clause, it is \emph{added} to the proof.  Sharing a clause
consists of passing a pointer and incrementing (atomically) the reference
count.  Dereferencing a clause (for instance during clause-database
reduction) decrements the reference count and when the reference counter
reaches zero the clause is deallocated and at the same time marked as
\emph{deleted} in the proof trace. In the meantime the clause is considered
alive.  As binary clauses are virtual and only occur in the watch lists,
they are handled as in other solvers with virtual binary clauses as long as
proof tracing is concerned.

All solver threads share the same proof trace file and writing
to that file is synchronized implicitly using the standard locking
mechanism of file I/O in \texttt{libc}.  In particular, the library makes
sure that calls to \texttt{fwrite} are executed atomically.  To avoid
additional locking, we first asynchronously collect complete proof lines in
thread local buffers before calling \texttt{fwrite} and then
rely on its implicit locking mechanism.

\section{Experiments on Proofs}
\label{sec:experiments}

We want to evaluate the
advantage of not copying the clauses in the proofs. But instead of implementing a
variant that really copies clauses, we \emph{fake copying}: the
clauses are \emph{only} duplicated in the proof, but not in the solver.
We argue that this approach gives similar performance as really copying
clauses, as it would be necessary in other multi-threaded solvers which
copy clauses, both in terms of proof size and checking time.

However, adapting inprocessing was a challenge, as our current version of
for instance bounded variable elimination during inprocessing requires that all
irredundant clauses are deduplicated.  Instead we \emph{deactivated}
any form of sequential inprocessing completely which requires deduplication.
Thus, in the following experiments we only report on a variant of \gimsatul with
initial preprocessing enabled, but only thread-local inprocessing enabled
(vivification and failed literal probing).

We consider the 96 unsatisfiable problems that are
solved by \gimsatul without inprocessing by all 1, 2, 4, 8, and 16-thread
configurations (due to non-determinism, rerunning benchmarks could lead to a
different set of solved problems though).  We use
unsatisfiable problems in order to be able to do backwards checking instead
of forward checking as is required if the problem is satisfiable or no
contradiction is derived. Due to the time required to check proofs
(\(\ge\)\SI{10}{\hour}), we were not able to run the benchmarks for 32
threads before the submission deadline either.

\begin{figure*}
  \centering
  \includegraphics[scale=0.6]{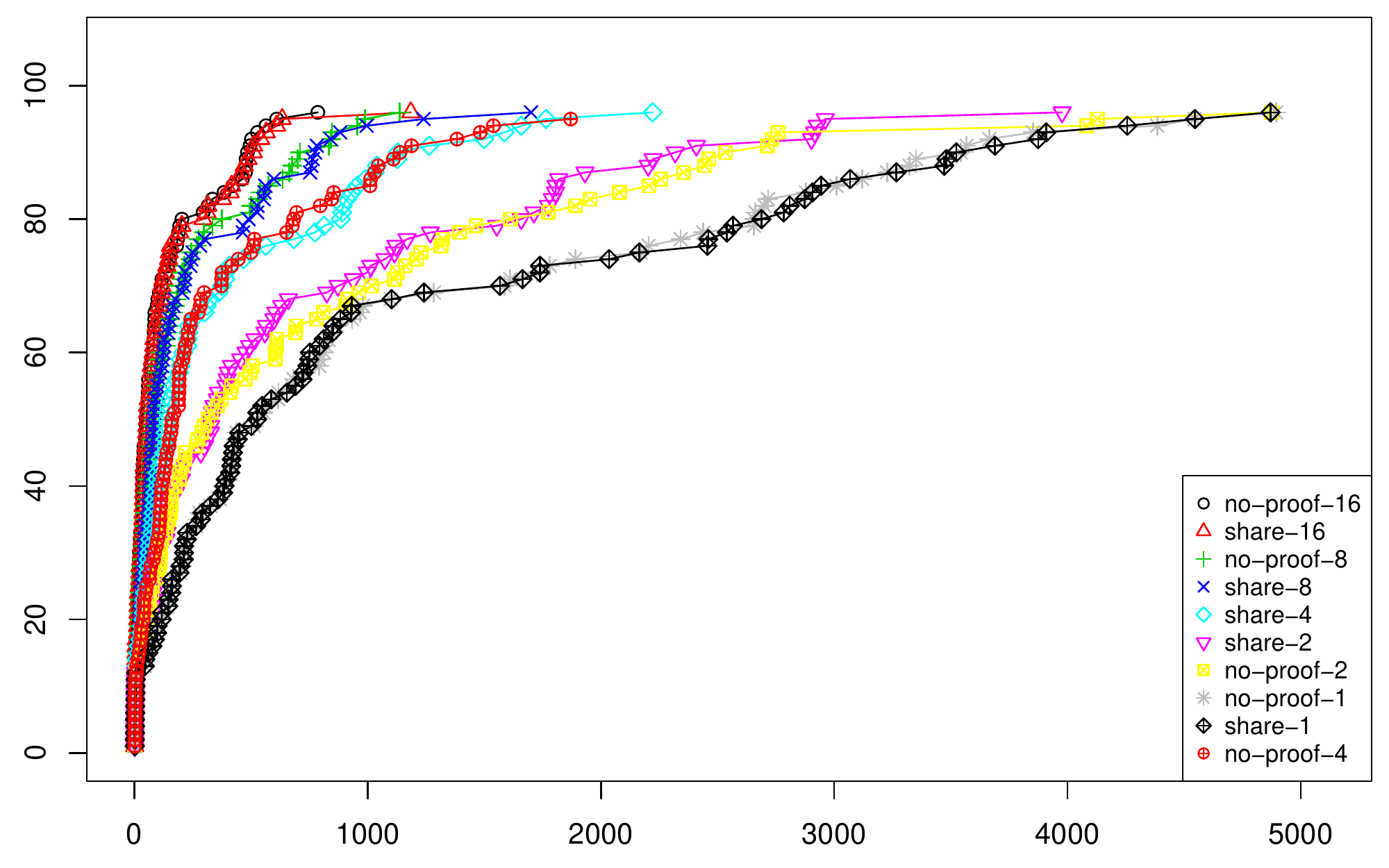}
  \caption{\small Cumulative distribution function (CDF) of the wall-clock solving
  time of \gimsatul with and without proof generation for the discussed variant with
  sequential inprocessing disabled}
  \label{fig:without-versus-with-proof-solving-times}
\end{figure*}

\PARAGRAPH{Does Proof Generation Cause a Slowdown?} Generating proofs is much
more costly in a context of parallel SAT solving because proof logging is
inherently single-threaded. However, according to our experiments in
Figure~\ref{fig:without-versus-with-proof-solving-times}, the cost of
proof generation is negligible.

\PARAGRAPH{Do Proofs become Longer or Shorter?} The answer to this question
amounts to answering: is the work done by other rings useful or are the lines
never used and could be removed. If the proofs are longer, then the work was
useless. If the proof have similar length, then the rings just do work that has
to be done too.

\begin{figure*}
  \centering
  \begin{subfigure}[b]{0.45\textwidth}
    \includegraphics[scale=0.6]{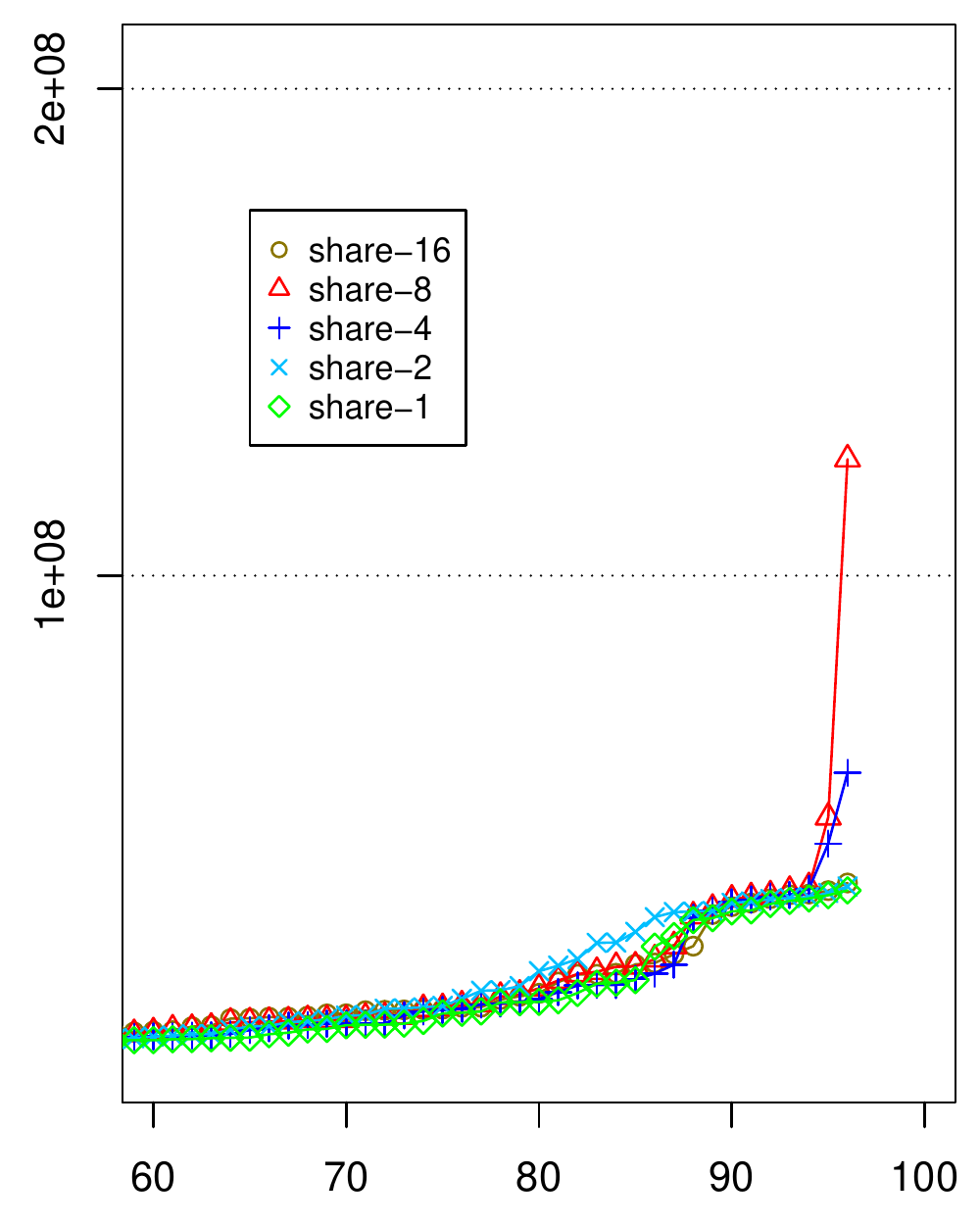}
    \caption{\small Proof length with sharing}
    \label{fig:proof-length-sharing-gimsatul}
  \end{subfigure}
  \quad
  \begin{subfigure}[b]{0.45\textwidth}
    \includegraphics[scale=0.6]{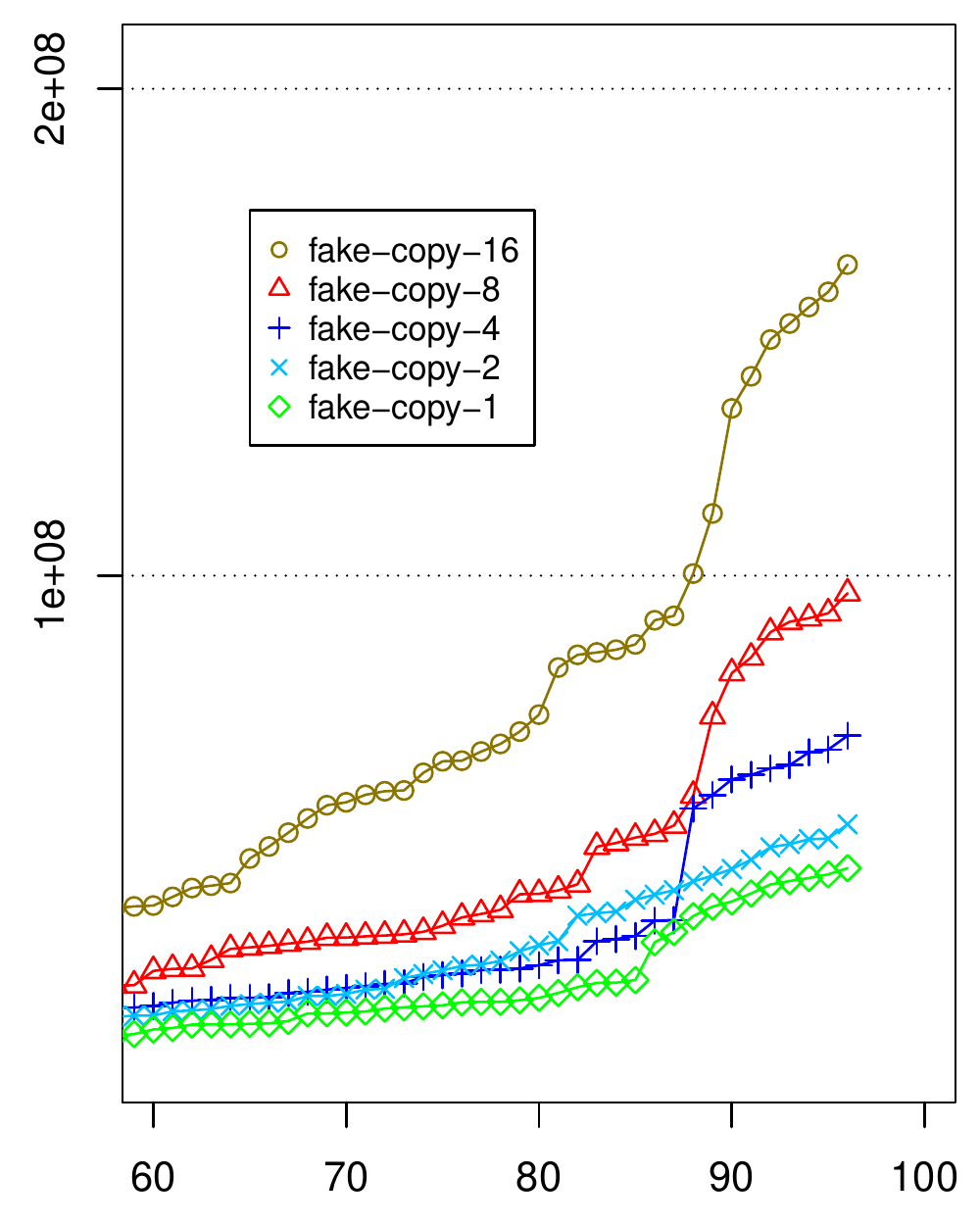}
    \caption{\small Proof length with fake copying}
    \label{fig:proof-length-fake-gimsatul}
  \end{subfigure}
  \caption{\small Proof length}
  \label{fig:proof-length-gimsatul}
\end{figure*}

In Figure~\ref{fig:proof-length-sharing-gimsatul} we show the proof size for
runs using different number of threads.  The plots are not completely
conclusive. The 2-thread
configuration produces larger proofs than the 1-thread and 4-thread versions.
We realized after looking at the results of the experiments that our portfolio
strategy might not be the best possible one and explain the bad behavior of 2 and 16
threads compared to 4 or 1. For 2 threads, instead of using one thread that focuses on SAT
(stable mode) and one that focuses on UNSAT (focused mode) following the idea of
Chanseok Oh~\cite{Oh-SAT-UNSAT}, one thread runs in focused mode and one in
alternating mode. For 16 threads, we have 8 threads that are running the same
strategy. Testing both assumptions is future work.

\PARAGRAPH{Is Sharing Clauses Useful?}
In Figure~\ref{fig:proof-length-fake-gimsatul}, we show the length of the proofs
without sharing the proofs. It is clear that the proofs are longer or even much
longer. Interestingly, for most problems, the size does not change much between
the 4-thread and the 2-thread version. This indicates that the number of
exchanged clauses is either limited or that the shared clauses are useful to
reduce the search space limiting the overhead: If every clause was shared and
entirely useless, the proofs would be \(n\) times as large for \(n\) threads.

\begin{figure*}
  \centering
  \begin{subfigure}[b]{0.45\textwidth}
    \includegraphics[scale=0.6]{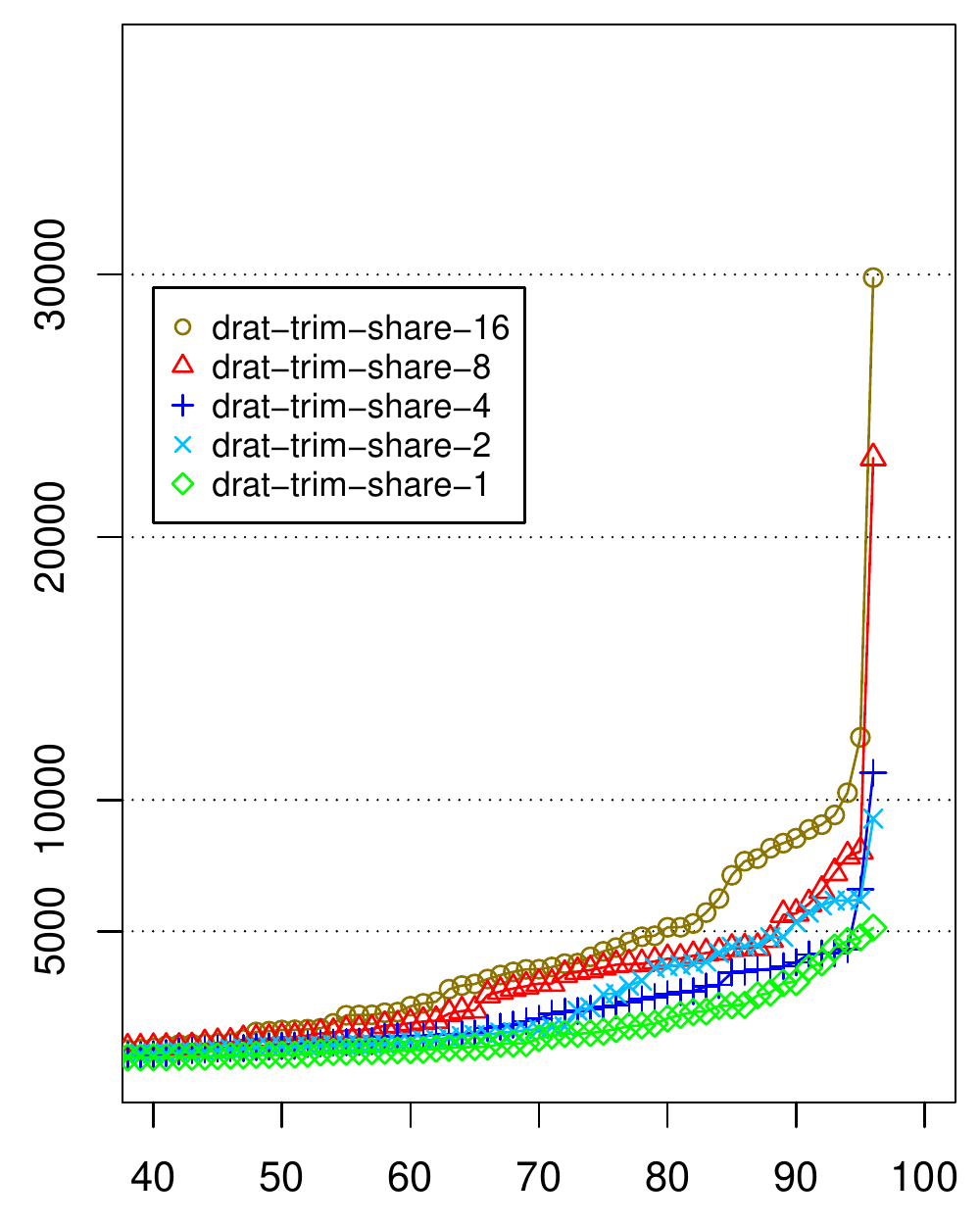}
    \caption{\small Proof checking with sharing}
    \label{fig:proof-checking-sharing-gimsatul}
  \end{subfigure}
  \quad
  \begin{subfigure}[b]{0.45\textwidth}
    \includegraphics[scale=0.6]{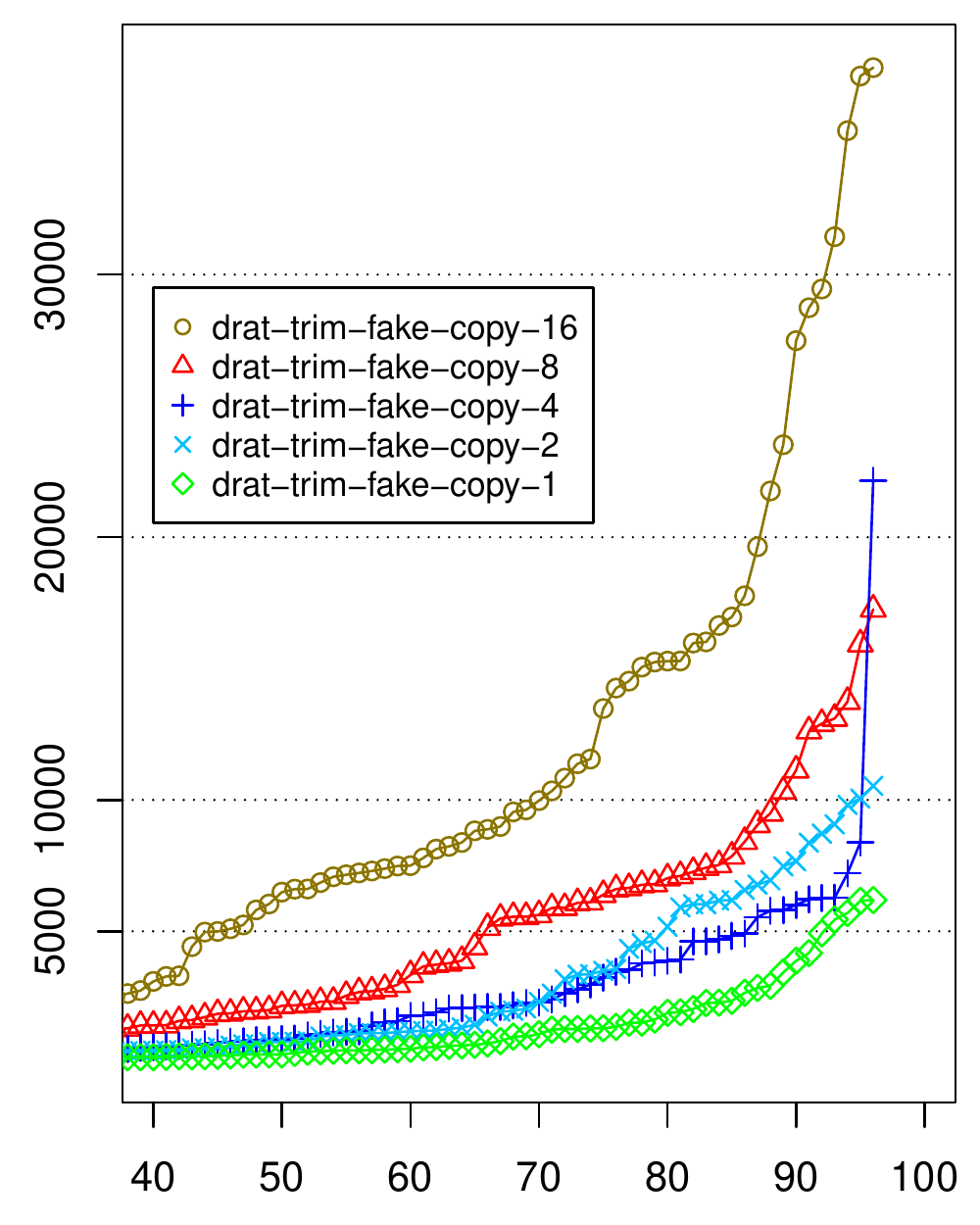}
    \caption{\small Proof checking with  fake copying}
    \label{fig:proof-checking-fake-gimsatul}
  \end{subfigure}
  \caption{\small Amount of time required to check the proofs produced by \gimsatul}
  \label{fig:proof-checking-gimsatul}
\end{figure*}

\PARAGRAPH{Is Proof Checking Easier?}
In Figure~\ref{fig:proof-checking-gimsatul}, we have plotted the amount of time
\tools{drat-trim} needed to check the proofs. It is important to notice that
\tools{drat-trim} does not seem to detect duplicates\footnote{Our
understanding of the
  source code of \tools{drat-trim} is that it uses a hash table only to count the number of clauses for
  deletions, but not to avoid relearning clauses.} and hence must reprove the lemma
each time and every copy must be propagated when either of them is propagating.

With sharing, checking scales better, but is still much slower
than in the single-threaded case. This can be explained by the fact that
\tools{drat-trim} does not know the context of learned clauses, i.e.,
which solver thread produced which clauses.
\iftrue
The checker has to treat them
all in the same way potentially increasing propagation across contexts,
which arguably yields an overhead during checking redundancy.
\fi

\PARAGRAPH{Should Proof Checking be added to the SAT Competition?}
This is the most controversial question without clear answer: even without
physical clause sharing, \tools{drat-trim} is able to check clauses. However,
even though checking 32-thread certificates is less than 32-times slower, it most
likely is too slow to be run in practice.  Thus we consider parallelizing DRAT
checkers as an important future work to solve this problem.
\iftrue
Alternatively
we will look into producing proof formats with antecedent information,
for which parallel checking is easier.
\fi

\section{Conclusion}
\label{sec:conclusion}

We presented the architecture of our newest solver \gimsatul which shares
clauses physically without copying.
Even though it only features a simple form of diversification
it scales linearly with the number of threads in our experiments.
We also study the
number of proof steps in this setting and observed that physical sharing yields smaller proofs.
Nearly all proofs generated by runs with multiple
solving threads have similar size as those produced
by a single thread.

We want to further explore
alternative clause exchange and search diversification strategies.  We might also
look into parallelizing variable elimination, subsumption and equivalent
literal substitution, which are currently run by a single thread, even
though this part does not seem to be a bottle-neck for large time-outs as
used in the SAT competition. Making the solver deterministic like ManyGlucose
would make the SAT solver easier to debug, and the biggest requirement, the time measurement by memory accesss, is already present in the code.

{\smallskip\small \PARAGRAPH{Acknowledgment.}  This work is supported by
  the Austrian Science Fund (FWF), NFN S11408-N23 (RiSE), and the LIT
  AI Lab funded by the State of Upper Austria.
  Some of the design choices for the clause sharing architecture
  of \gimsatul are inspired by discussions with
  Christoph Kirsch and his former students Martin Aigner and Andreas Haas
  and other colleagues in RiSE~\cite{kstack}.
  The idea of using separate distributed pools inspired
  by $k$-stacks, in order to avoid a global lock-less queue and at the same time reduce
  false sharing was first proposed to us by Andreas in 2012 during our joint
  project.  We also thank Sonja Gurtner for many timely textual improvements.
  }

\bibliographystyle{splncs04}
\bibliography{paper}

\end{document}